\documentclass[twocolumn,english,aps,prb,floats]{revtex4-2}
\usepackage[T1]{fontenc}
\usepackage[latin9]{inputenc}
\usepackage{color}
\usepackage{babel}
\usepackage{amsmath}
\usepackage{amssymb}
\usepackage{graphicx}
\usepackage{esint}
\usepackage[unicode=true,
 bookmarks=false,
 breaklinks=false,pdfborder={0 0 1},backref=section,colorlinks=false]
 {hyperref}

\makeatletter

\newcommand{\lyxdot}{.}

\usepackage{babel}

\@ifundefined{textcolor}{}{%
 \definecolor{BLACK}{gray}{0}
 \definecolor{WHITE}{gray}{1}
 \definecolor{RED}{rgb}{1,0,0}
 \definecolor{GREEN}{rgb}{0,1,0}
 \definecolor{BLUE}{rgb}{0,0,1}
 \definecolor{CYAN}{cmyk}{1,0,0,0}
 \definecolor{MAGENTA}{cmyk}{0,1,0,0}
 \definecolor{YELLOW}{cmyk}{0,0,1,0}
}

\usepackage{ifpdf}\usepackage{bm}

\makeatother

\begin{document}
\title{Melting and Freezing of a Skyrmion Lattice}
\author{Dmitry A. Garanin, Jorge F. Soriano, and Eugene M. Chudnovsky}
\affiliation{Physics Department, Herbert H. Lehman College and Graduate School,
The City University of New York, 250 Bedford Park Boulevard West,
Bronx, New York 10468-1589, USA }
\date{\today}
\begin{abstract}
We report comprehensive Monte-Carlo studies of the melting of skyrmion
lattices in systems of small, medium, and large sizes with the number
of skyrmions ranging from $10^{3}$ to over $10^{5}$. Large systems
exhibit hysteresis similar to that observed in real experiments on
the melting of skyrmion lattices. For sufficiently small systems which
achieve thermal equilibrium, a fully reversible sharp solid-liquid
transition on temperature with no intermediate hexatic phase is observed.
A similar behavior is found on changing the magnetic field that provides
the control of pressure in the skyrmion lattice. We find that on heating
the melting transition occurs via a formation of grains with different
orientations of hexagonal axes. On cooling, the fluctuating grains
coalesce into larger clusters until a uniform orientation of hexagonal
axes is slowly established. The observed scenario is caused by collective
effects involving defects and is more complex than a simple picture
of a transition driven by the unbinding and annihilation of dislocation
and disclination pairs. 
\end{abstract}
\maketitle

\section{Introduction}

Recently, several groups have reported theoretical and experimental
studies of the melting of skyrmion lattices (SkL) in ferromagnetic
films \citep{Ambrose,Nishikawa-PRB2019,Zazvorka2020,Huang-Nat2020,Balaz-PRB2021,McCray2022,GC-PRB2023,Meisenheimer2023}.
They have been driven by the interest to the fundamental problem of
crystallinity and melting transition in two dimensions (2D), which
generated controversies for almost a century of theoretical and experimental
research, see, e.g., Refs.\ \citep{Kosterlitz-Review2016,Ryzhov2023}
for review. This paper contains a detailed study of the melting of
the skyrmion lattice. The brief history of the subject is outlined
below.

The study of 2D crystals began with the work of Peierls \citep{Peierls}
who had shown that in 2D the average square deviation of atoms $\langle{\bf u}^{2}\rangle$
from equilibrium positions in a crystal diverges with the size of
the system $L$ as $\ln L$ due to thermal fluctuations. Thirty years
later, Mermin and Wagner \citep{Mermin}, and independently Hohenberg,
\citep{Hohenberg} argued that the long-range order cannot exist in
2D not only for crystals but also for magnets (no net magnetization)
and superfluids (no Bose condensation). This, however, contradicted
the numerical work on hard disks \citep{Adler1962} performed in the
early 1960s, which showed a melting transition between solid and liquid
states. Also, contrary to the theory, the superfluidity in 2D helium
films was demonstrated experimentally in the late 1960s \citep{Kagiwada}.
It was subsequently realized that while fluctuations indeed grew in
2D as $\ln L$, they did not prevent the system from developing rigidity
at low temperatures, which for a 2D crystal meant finite elastic moduli.

In 1971 Berezinskii demonstrated theoretically that the destruction
of the long-range order in thin magnetic films and 2D crystals should
occur through the creation of topological defects \citep{Berezinskii}.
One year later, Kosterlitz and Thouless (KT) \cite{KT} arrived at
the same conclusion for 2D solids and superfluids. The KT results
for the temperature dependence of the superfluid density in a helium
film were confirmed by Bishop and Reppy in the late 1970s \citep{Bishop}.
Soon, however, Halperin and Nelson \citep{HN-PRL1978,NH-PRB1979},
and independently Young \citep{Young-PRB1979}, realized that as far
as the solids were concerned the KT theory was missing the angular
terms in the energy of interaction between dislocations. By including
such terms in the Hamiltonian, they showed that the melting of a 2D
crystal occurred into an anisotropic liquid with the exponential decay
of translational correlations but only the algebraic decay of orientational
correlations, which they called a hexatic liquid. The latter, according
to the KTHNY theory, melted into the isotropic liquid due to the creation
of disclinations via a second transition at a higher temperature.

In contrast with the KTHNY theory, however, early simulations that
used molecular dynamics, see, e.g., Ref.\ \citep{Broughton1982},
emphasized the importance of short-wave phonons that can drive a conventional
first-order melting in 2D as they do in 3D via the Lindemann melting
criterion \citep{Lindenmann}. (It was recently suggested \cite{Khrapak}
that in 2D the Lindemann criterion essentially coincides with the
KTHNY condition for spontaneous creation of dislocations). In addition,
in the early 1980s, Chui \cite{Chui} and independently Saito \cite{Saito},
considering the 2D melting problem in terms of interacting dislocations,
argued that the KTHNY scenario of melting via the spontaneous creation
of independent dislocation pairs by thermal agitation was self-consistent
only at a sufficiently large energy of the dislocation core, compared
to their interaction energy. As the core energy is lowered, dislocations
are created in clusters. Close to the transition temperature they
aggregate into the grain boundaries that break the 2D crystal into
differently oriented grains. Such a possibility was first acknowledged
in the earlier studies of the melting of a 2D electron crystal \citep{Fisher1979}.
The energy of the dislocation core is difficult to measure directly
and it requires a computation at the level of individual atoms. The
lack of that knowledge creates the biggest uncertainty about the nature
of the melting transition in a 2D system.

Numerous experimental studies performed on 2D crystals in the 1980s
and 1990s, see e.g. Ref.\ \citep{Janke1988}, showed a first-order
melting transition, as in 3D, with no sign of the hexatic phase. The
existence of the latter at low temperatures due to disorder was predicted
\citep{EC-solid,EC-SC} and observed in vortex lattices of type-II
superconductors \citep{Murray}, but convincing evidence of the thermally
induced KTHNY two-step melting emerged from the study of lattices
of colloidal particles \citep{Grunberg,Zanghellini} only in the 2000s.
Subsequently, it was established by numerous studies that the melting
scenario was not universal. Besides the energy of the dislocation
core, it depends on the interaction potential, the shape of the interacting
particles, and the symmetry of the crystal lattice, see, e.g., Refs.\ \citep{Dietel2006,Kapfer2015,Anderson2017,Tsiok2022}
and references therein.

The interest in skyrmions in magnetic films, besides the beauty of
the physics \cite{belpol75jetpl} and imagery associated with them,
has been driven by the prospect of developing topologically protected
information technology \citep{Nagaosa2013,Zhang2015,Klaui2016,Hoffmann-PhysRep2017,Fert-Nature2017}.
In materials lacking inversion symmetry, individual skyrmions are
stabilized by Dzyaloshinskii-Moriya interaction (DMI) \citep{Bogdanov1989,Bogdanov94,Bogdanov-Nature2006,Heinze-Nature2011,Boulle-NatNano2016,Leonov-NJP2016}.
Other mechanisms of stabilization of skyrmions include frustrated
exchange interactions \citep{Leonov-NatCom2015,Zhang-NatCom2017},
magnetic anisotropy \citep{IvanovPRB06,Lin-PRB2016}, disorder \citep{CG-NJP2018},
and geometrical confinement \citep{Moutafis-PRB2009}.

Theoretical research on 2D lattices of magnetic vortices in systems
with Dzyaloshinskii-Moriya (DMI) interaction was pioneered by Bogdanov
and Hubert \citep{Bogdanov94}. They obtained the magnetic phase diagram
separating the conventional laminar domain structure from the magnetic
state with a periodic arrangement of vortices. The first evidence
of regular skyrmion lattices was reported in FeCoSi films by Yu et
al. \citep{Yu2010} with the help of the real-space Lorentz transmission
electron microscopy. It was followed by an avalanche of papers demonstrating
transitions between uniformly magnetized states, laminar domains,
and skyrmion crystals on the magnetic field and temperature, see,
e.g., Refs. \citep{GCZZ-MMM2021,Dohi-ARCMP2022} and references therein.

\begin{figure}
\centering{}\includegraphics[width=8cm]{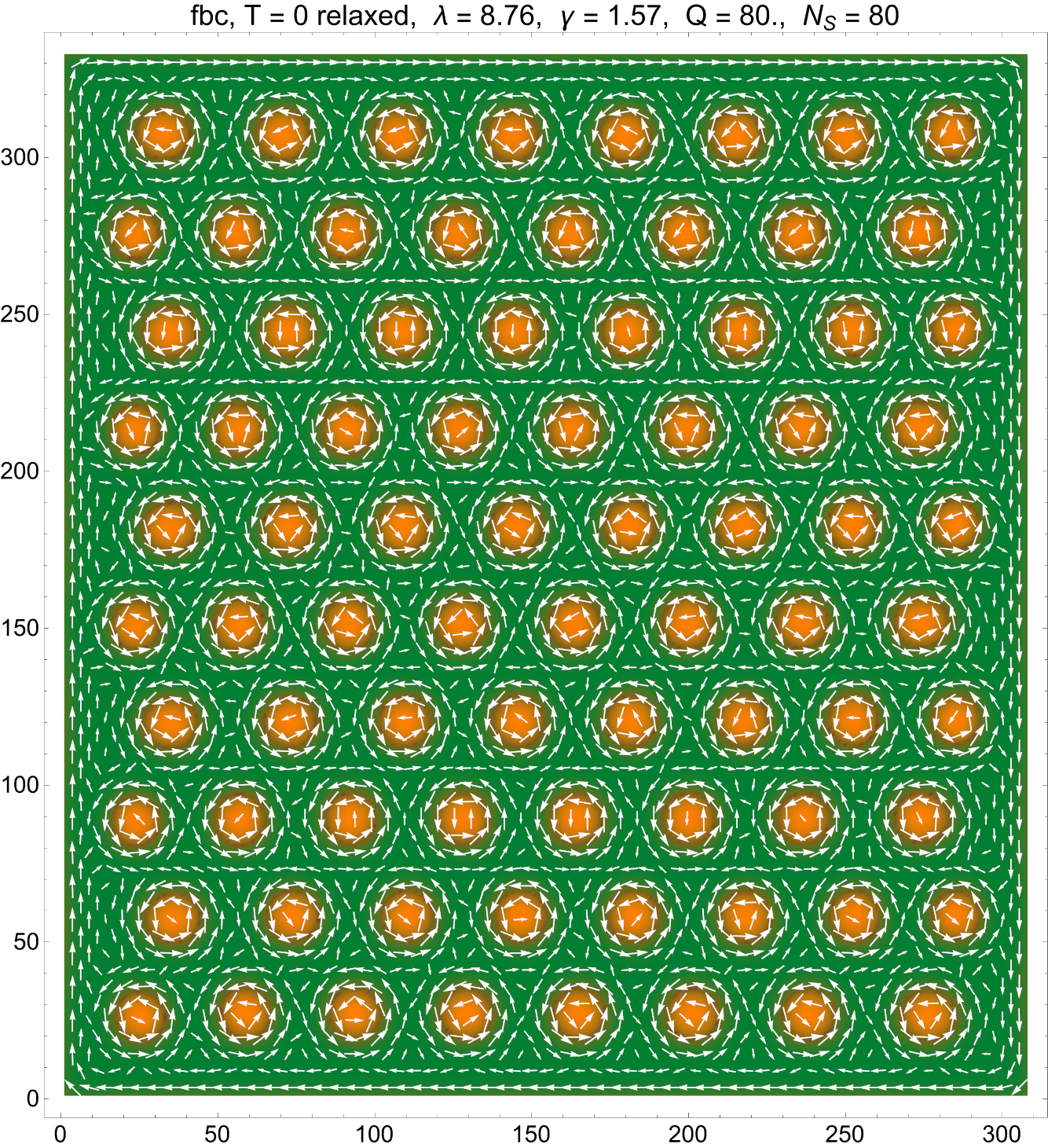}
\caption{Triangular skyrmion lattice in the system with free boundary condition
at $T=0$. Spin components are color-coded, $s_{z}=-1$ green, $s_{z}=1$
orange. White arrows show the in-plane spin components $s_{x}$ and
$s_{y}$.}
\label{Fig-SkL} 
\end{figure}

In recent years it was realized that the skyrmion lattice (see Fig.\ \ref{Fig-SkL})
provides another rich system for the study of 2D melting. An advantage
here is that the melting transition can be achieved on both temperature
and the magnetic field. Conflicting results have been obtained so
far by theorists and experimentalists. Nishikawa et al. \citep{Nishikawa-PRB2019}
used Monte Carlo simulations for the lattice-spins model to show a
direct melting into an isotropic 2D liquid, with no evidence of the
intermediate hexatic phase. Monte-Carlo studies \citep{GC-PRB2023}
of lattices of up to $10^{5}$ skyrmions considered as point particles
with repulsive interaction obtained from a microscopic spin model
\citep{CGC-JPCM2020} rendered a similar conclusion. However, measurements
of Huang et al. \citep{Huang-Nat2020} performed on a nano-slab of
Cu$_{2}$OSeO$_{3}$ by using cryo-Lorentz transmission electron microscopy
reported a two-step KTHNY melting transition. Such a transition was
also observed in the simulations of skyrmion lattices in a GaV$_{4}$S$_{8}$
spinel by Baláž et al. \citep{Balaz-PRB2021}.

Nonequilibrium effects hinder comparison with the analytical theory
of the 2D melting that assumes thermal equilibrium. For example, thermal
studies of skyrmion lattices in a stack of Ta(5)/Co$_{20}$Fe$_{60}$B$_{20}$(0.9)/Ta(0.08)/MgO(2)/Ta(5)
by Zavorka et al. \citep{Zazvorka2020}, while providing good data
on the orientational correlations, did not produce consistent results
on the nature of the phase transition due to long equilibration times.
McGray et al. \citep{McCray2022} reported hysteresis in the temperature
dependence of the orientational order in the skyrmion lattices of
Fe$_{3}$GeTe$_{2}$ similar to that in the field-cooled/zero-field-cooled
(FC/ZFC) magnetization studies of the arrays of single-domain magnetic
particles in a solid matrix \citep{Lectures}. According to Meisenheimer
et al. \citep{Meisenheimer2023}, disorder and pinning of skyrmions
inhibited simple scenarios of the melting transition in a polar van
der Waals magnet (Fe$_{0.5}$Co$_{0.5}$)$_{5}$GeTe$_{2}$.

Such a broad range of melting scenarios in the skyrmion lattices can
be explained by several factors. One factor can be the difference
in the dislocation core energy for different materials. The slow equilibration
in both experiments and numerical work is another factor. We try to
overcome the latter problem in our Monte Carlo simulations while also
elucidating nonequilibrium effects that can be seen in real experiments.
Ambrose and Stamps \citep{Ambrose} mentioned the distortion of the
skyrmion profile on approaching the melting transition, which they
observed in the Monte Carlo studies that used parameters of FeCoSi
and FeGe. It was confirmed that such effect does exist at the melting
transition where it it responsible for the kink in the temperature
dependence of the magnetization \citep{Garanin2024}. Since the number
of magnetic systems in which skyrmion lattices have been observed
experimentally increases rapidly, in our studies we stuck to the most
common range of parameters without specifying a concrete material.
The goal is to elucidate features of the skyrmion matter that may
be seen in the melting and freezing of other 2D systems.

Apart from elucidating the thermodynamics of the melting/freezing
transition, we investigate dislocations which spontaneously emerge
when a 2D skyrmion solid is heated. The picture that we observe in
Monte Carlo simulations and illustrate with images in the paper and
with movies in the supplemental material, clearly points toward the
Chui-Saito melting scenario \citep{Chui, Saito} in which the dislocations
emerge in clusters on approaching the melting transition. At the transition,
they aggregate into grain boundaries that break the 2D solid with
a quasi-long-range orientational order into a system of fluctuating
differently oriented grains. The transition is well defined by the
abrupt change in the orientational order parameter. However, the orientationally
ordered grains of fluctuating shape and size are also seen in the
liquid phase. We study their properties with an eye on real experiments
that can image the structure of the skyrmion matter near the melting/freezing
transition.

The paper is organized as follows. The model of the skyrmion lattice,
the order parameter, and the relevant thermodynamics are explained
in Sec.\ \ref{Sec_Model}. The numerical method is described in Sec.\ \ref{Sec_Numerical}.
Thermodynamic quantities, including the vicinity of the melting/freezing
transition, are computed in Sec.\ \ref{Sec_Thermodynamics}. The
characteristics of the defects that spontaneously emerge on rising
temperature are computed and illustrated by images in Sec.\ \ref{Sec_Defects}.
The nonequilibrium effects are studied in Sec.\ \ref{Sec_Nonequilbrium}.
Our conclusions and suggestions for experiments are presented in Sec.\ \ref{Sec_Conclusions}.

\section{The model}

\label{Sec_Model}

\subsection{Skyrmions in the lattice-spins model}

We consider skyrmions in the model of ferromagnetically coupled three-component
classical spin vectors ${\bf s}_{i}$ of length $1$ on a square lattice
with the energy given by 
\begin{eqnarray}
\mathcal{H} & = & -\frac{1}{2}\sum_{ij}J_{ij}\mathbf{s}_{i}\cdot\mathbf{s}_{j}-H\sum_{i}s_{iz}\nonumber \\
 &  & -A\sum_{i}\left[(\mathbf{s}_{i}\times\mathbf{s}_{i+\delta_{x}})_{x}+(\mathbf{s}_{i}\times\mathbf{s}_{i+\delta_{y}})_{y}\right].\label{Hamiltonian}
\end{eqnarray}
The nearest neighbors exchange interaction of strength $J>0$ favors
ferromagnetic ordering and incorporates the actual length of the spin.
The stabilizing field $H$ in the second (Zeeman) term is applied
in the negative \textit{z}-direction, $H<0$. The third term in Eq.\ (\ref{Hamiltonian})
is the Dzyaloshinskii-Moriya interaction (DMI) of the Bloch type of
strength $A$, with $\delta_{x}$ and $\delta_{y}$ standing for to
the nearest lattice site in the positive $x$ or $y$ direction. In
this configuration, the dominant direction of spins is down and that
in the skyrmions is up. The DMI of the Néel type is described by the
term $(\mathbf{s}_{i}\times\mathbf{s}_{i+\delta_{x}})_{y}-(\mathbf{s}_{i}\times\mathbf{s}_{i+\delta_{y}})_{x}$
and leads to similar results.

Two-dimensional configurations of the spin field are characterized
by the topological charge \cite{belpol75jetpl}
\begin{equation}
Q=\int\frac{dxdy}{4\pi}\:{\bf s}\cdot\frac{\partial{\bf s}}{\partial x}\times\frac{\partial{\bf s}}{\partial y}\label{Q}
\end{equation}
that takes quantized values $Q=0,\pm1,\pm2,...$. For the pure-exchange
model, the analytical solution for topological configurations with
a given value of $Q$ was found by Belavin and Polyakov \cite{belpol75jetpl}
(see also Ref. \cite{CGC-JPCM2020} for a more transparent approach).
Due to the scale invariance of the exchange interaction in 2D, the
energy of a skyrmion or antiskyrmion ($Q=\pm1$) with respect to the
uniform state, $\Delta E_{\mathrm{BP}}=4\pi J$, does not depend on
their size $\lambda$. In a continuous spin-field model, the conservation
of the topological charge prevents skyrmions from decaying. Finite
lattice spacing $a$ breaks this invariance by adding a term of the
order $-\left(a/\lambda\right)^{2}$ to the energy, which leads to
the skyrmion collapse \citep{CCG-PRB2012}. This result was generalized
for topological structures with any $Q$ in Ref. \cite{capgarchu19prr}.

In the presence of the DMI, only skyrmions, but not antiskyrmions,
exist. To find a single-skyrmion spin configuration numerically, one
can start with any state with $Q=1$ and perform energy minimization.
The energy of a single skyrmion with respect to the uniform ferromagnetic
state becomes negative for $A$ large enough and $|H|$ small enough,
making the uniform state unstable to the creation of skyrmions. However,
skyrmions compete with the laminar domains. A stable skyrmion solution
exists in the field interval $H_{s}\leq|H|\leq H_{c}$. Below $H_{s}\simeq0.55A^{2}/J$
skyrmions become unstable against converting into a laminar domain
structure \citep{GC-PRB2023}. Above $H_{c}\simeq0.97A^{3/2}/J^{1/2}$
skyrmions collapse \citep{derchugar2018}. The ratio of these fields
is $H_{c}/H_{s}\simeq1.76\left(J/A\right)^{1/2}$. In this paper,
we use the value $A/J=0.2$ which is typical for materials with nanoscale
skyrmions, see, e.g.\textcolor{blue}{{} \cite{Leonov-NJP2016,Camley-2023}}.
In this case, $H_{c}/J=0.034$ and $H_{s}/J=0.022$. Figure \ref{Fig-SkL}
shows a Bloch-skyrmion lattice in our model, Eq.(\ref{Hamiltonian}),
obtained by the numerical energy minimization at $T=0$ for $A/J=0.2$
and $H/J=-0.025$.

\subsection{Skyrmions as point particles}

There is a short-range repulsion between skyrmions due to DMI \cite{CGC-JPCM2020}
\begin{equation}
U(r)\simeq F\exp\left(-\frac{r}{\delta_{H}}\right),\qquad F\equiv60J\left(\frac{A^{2}}{JH}\right)^{2},\label{U_interaction}
\end{equation}
where $r$ is the distance between the skyrmions' centers and $\delta_{H}=a\sqrt{J/|H|}$
is the magnetic length. In sufficiently dense skyrmion lattices that
we study here, this interaction dominates, whereas dipole-dipole repulsion
of skyrmions can be neglected. This result allows considering a skyrmion
lattice as a system of point particles with a repulsive interaction,
Eq. (\ref{U_interaction}), that is computationally easier than using
the full model of lattice spins, Eq. (\ref{Hamiltonian}).

The number of skyrmions $N_{S}$ in the system at $T=0$ can be found
by the minimization of the total energy. For a triangular lattice
of skyrmions of the period $a_{S}$ the unit-cell area is $a_{S}^{2}\sqrt{3}/2$
and $N_{S}$ can be expressed as 
\begin{equation}
N_{S}=\frac{2}{\sqrt{3}}\frac{S}{a_{S}^{2}},\label{NS_via_aS}
\end{equation}
where $S$ is the total area of the system. Each skyrmion in the lattice
interacts with its six nearest neighbors, whereas interactions with
further neighbors can be neglected. The energy per skyrmion in the
perfect lattice is 
\begin{equation}
E_{0}=\varDelta E+3F\exp\left(-\frac{a_{S}}{\delta_{H}}\right),\label{E0_def}
\end{equation}
where $\Delta E<0$ is the skyrmion's core energy computed from the
lattice-spins model \cite{GC-PRB2023}. The equilibrium state at $T=0$
can be obtained by the minimization of the total energy, $\mathcal{E}=N_{S}E_{0}$
with respect to $a_{S}$, as described in Ref.\ \citep{GC-PRB2023}.
For $A/J=0.2$ and $H/J=-0.025$ one has $\Delta E/J=-4.23$, $\delta_{H}=6.32a$,
and $F/J=154$ that results in $a_{S}=38.5a$. The interaction energy
between nearest neighbors $U_{0}=F\exp\left(-a_{S}/\delta_{H}\right)=0.349J$
defines the scale for the skyrmion-lattice melting temperature $T_{m}/J\simeq0.12$. 

In the temperature range below as well as above melting the number
of skyrmions in the system does not change because both annihilating
a skyrmion or adding a skyrmion to the system requires overcoming
large energy barriers \cite{GC-PRB2023}. In the processes under investigation
$\Delta E$ plays no role. One can consider metastable states in which
$N_{S}$ deviates from the value found by the energy minimization.

The quantity describing the orientation of the hexagon of the nearest
neighbors of any skyrmion $i$ in a 2D lattice is the local hexagonality
\begin{equation}
\Psi_{i}=\frac{1}{6}\sum_{j}\exp(6i\theta_{ij}).\label{Hexagonality_def}
\end{equation}
Here the summation is carried out over the six nearest neighbors $j$
of the skyrmion $i$ (without the distance cut-off), $\theta_{ij}$
is the angle between the $ij$ bond and any fixed direction in the
lattice. If $\theta$ is counted from the direction of the $x$-axis
(that is our choice), and two of the sides of the perfect hexagon
coincide with the $x$-axis (as in the figures above), all terms in
the sum are equal to $1$, and thus $\Psi_{i}=1$. We call this \textit{horizontal}
orientation of hexagons and label it by the red color. Rotation of
the hexagons by 30$^{\circ}$ from the horizontal orientation results
in the \textit{vertical} orientation for which $\Psi_{i}=-1$. For
any other orientation of a perfect hexagon, $\Psi_{i}$ is a complex
number of modulus 1. The angle $\phi_{i}$ by which the hexagon $i$
is rotated from its initial horizontal orientation is related to the
phase angle $\Theta_{i}$ in $\Psi_{i}=\left|\Psi_{i}\right|e^{i\Theta_{i}}$
as $\phi_{i}=\Theta_{i}/6$.

At finite temperatures, orientations of the bonds fluctuate and the
condition $|\Psi_{i}|=1$ no longer holds. The quality of hexagons
can be described by global hexagonality value defined as
\begin{equation}
V_{6}\equiv\sqrt{\frac{1}{N_{S}}\sum_{i}|\Psi_{i}|^{2}}.\label{V6_def}
\end{equation}
At temperatures well above the melting transition, when even the short-range
order is completely destroyed, the orientations of the bonds and the
angles $\theta_{ij}$ become random. In this limit $V_{6}=\sqrt{1/6}$
as each particle has six nearest neighbors on average. The common
orientation of hexagons in the lattice is described by the complex
order parameter which is defined as local hexagonality averaged over
the system:
\begin{equation}
\Psi=\frac{1}{N_{S}}\sum_{i}\Psi_{i}.\label{Psi_def}
\end{equation}

We compute the averages of these quantities over the Monte Carlo process
obtaining 
\begin{equation}
V_{6}\equiv\left\langle \sqrt{\frac{1}{N_{S}}\sum_{i}|\Psi_{i}|^{2}}\right\rangle ,\qquad O_{6}=\left\langle \Psi\right\rangle .\label{O6_def}
\end{equation}
In addition, we define 
\begin{equation}
G_{6}=\sqrt{\left\langle \left|\Psi\right|^{2}\right\rangle }.\label{G6_def}
\end{equation}
For the hexagonality value one has $V_{6}^{2}>0.5$ in the solid phase
and $V_{6}^{2}<0.5$ in the liquid phase. In the solid phase $G_{6}\cong\left|O_{6}\right|$.
In the liquid phase $O_{6}$ reduces to zero by a long thermalization,
whereas $G_{6}$ remains finite and scales with $1/\sqrt{N_{S}}$.
One can formally introduce the susceptibility
\begin{equation}
\chi=\frac{N_{S}}{3T}\left(\left\langle \left|\Psi\right|^{2}\right\rangle -\left|\left\langle \Psi\right\rangle \right|^{2}\right)\label{chi_def}
\end{equation}
that contains both $O_{6}$ and $G_{6}$ and peaks at the melting/freezing
point.

Similarly, one can compute the heat capacity via energy fluctuations:
\begin{equation}
C=\frac{d\left\langle E\right\rangle }{dT}=\frac{N_{S}}{T}\left(\left\langle E^{2}\right\rangle -\left\langle E\right\rangle ^{2}\right)\label{C_def}
\end{equation}
with the energy per skyrmion given by
\begin{equation}
E=\Delta E+\frac{F}{2N_{S}}\sum_{ij}e^{-r_{ij}/\delta_{H}},\label{E_def-1}
\end{equation}
where $\Delta E$ does not contribute.

The quantity $G_{6}$ also provides the information about orientational
correlations in the system:
\begin{equation}
G_{6}^{2}=\left\langle \frac{1}{N_{S}^{2}}\sum_{ij}\Psi_{i}\Psi_{j}^{*}\right\rangle =\frac{V_{6}^{2}}{N_{S}}\sum_{i-j}C_{6,i-j},
\end{equation}
where
\begin{equation}
C_{6,ij}=\frac{1}{V_{6}^{2}}\left\langle \Psi_{i}\Psi_{j}^{*}\right\rangle .
\end{equation}
is the correlation function of the hexagonalities. Here the factor
$V_{6}^{2}$ is introduced to account for distortions of hexagons
at nonzero temperatures and enforce $C_{6,ii}=1$. If correlations
are long, one can use continuous approximation
\begin{equation}
G_{6}^{2}=\frac{V_{6}^{2}}{N_{S}}\int\frac{d^{2}r}{a_{S}^{2}\sqrt{3}/2}C_{6}(r),
\end{equation}
see Eq. (\ref{NS_via_aS}). For $C_{6}(r)=\exp\left[-\left(r/R_{6}\right)^{p}\right]$
one obtains
\begin{equation}
G_{6}^{2}=\pi K_{p}\frac{2V_{6}^{2}R_{6}^{2}}{N_{S}\sqrt{3}a_{S}^{2}}\Longrightarrow\frac{R_{6}}{a_{S}}=\frac{G_{6}}{V_{6}}\sqrt{\frac{\sqrt{3}N_{S}}{2\pi K_{p}}},\label{R6}
\end{equation}
where $K_{1}=2$ and $K_{2}=1$. Theory of critical phenomena considers
correlation functions of the type $C(r)\propto r^{2-d-\eta}e^{-r/R}$,
where $d$ is the dimensionality of the space. With this Ansatz, one
obtains $G_{6}^{2}\propto R_{6}^{2-\eta}$ and further the scaling
relation $\gamma=\nu(2-\eta)$. The critical index $\eta$ could be
extracted from the behavior of the correlation function at criticality.
However, in our case, there is a very steep temperature dependence
at the melting/freezing point consistent with only a zero value of
the critical index $\beta$ for the orientational order parameter.
Thus the state of the system at criticality is difficult to define
and we are not using the critical Ansatz for the orientational CF.

\section{Numerical method}

\label{Sec_Numerical}

Throughout the paper, we consider the system with periodic boundary
conditions (pbc) and sizes $L_{x}\cong L_{y}$ compatible with the
lattice without distortions (see Ref. \cite{GC-PRB2023} for details).
We studied the small system having the linear size 30 in the SkL units
$a_{S}$ and containing $N_{S}=1080$ skyrmions, the medium system
with the linear size 100 and 1$1600$ skyrmions, and the large system
with the linear size 300 and 104400 skyrmions.

To equilibrate the system at a particular temperature and a magnetic
field, we used the classical Metropolis Monte Carlo (MC) method. Particles
were successively moved by the random vector $\Delta\mathbf{r}$ of
the length $\left|\Delta\mathbf{r}\right|=\mathrm{dispmax\times rand}$,
where rand is a random real number in the range $(0,1)$ and dispmax
is the maximal displacement chosen automatically to maximize the effective
displacement $\mathrm{dispeff}=0.5\mathrm{dispmax}\times\mathrm{acceptancerate}$.
It turns out that near and above melting the maximum of dispeff on
dispmax is very flat and there is a tendency for dispmax to become
comparable with $a_{S}$. Because of that, we introduced a limitation
$\mathrm{dispmax}<\mathrm{dispmaxmax}\equiv0.3a_{S}$. After each
trial move, the energy change $\Delta E_{\mathrm{trial}}$ was computed
and the move was accepted if $\exp\left(-\Delta E_{\mathrm{trial}}/T\right)>\mathrm{rand}$.

To properly deal with the interacting neighbors, we introduced the
interaction cutoff $r_{\mathrm{cutoff}}=0.95\sqrt{3}a_{S}$ that is
just shy of the next-nearest distance in the triangular lattice, $\sqrt{3}a_{S}$,
and subdivided the system into square bins of the size $1.05r_{\mathrm{cutoff}}+\mathrm{dispmaxmax}$.
For any skyrmion, interacting neighbors are looked up in the same
bin and in the eight surrounding bins. This drastically reduces the
procedure of finding the interacting neighbors that becomes not straightforward
in the liquid phase where the particles are moving by large distances
and the interacting neighbors change. The fast binning procedure was
repeated after each Monte Carlo step (MCS), that is, after updating
all particles in the system.

\begin{figure}
\begin{centering}
\includegraphics[width=8cm]{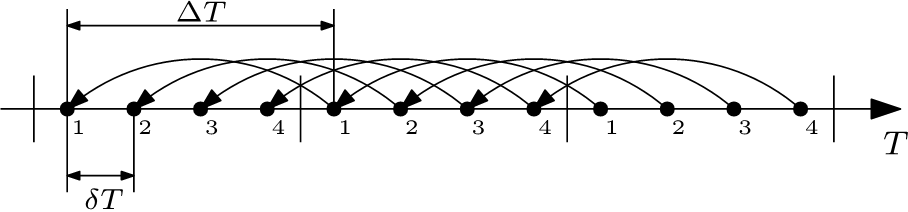}
\par\end{centering}
\caption{Parallelized relay computation -- a four-core example with lowering
temperature.}

\label{Fig-Relay_computation}
\end{figure}

\begin{figure}
\centering{}\includegraphics[width=8cm]{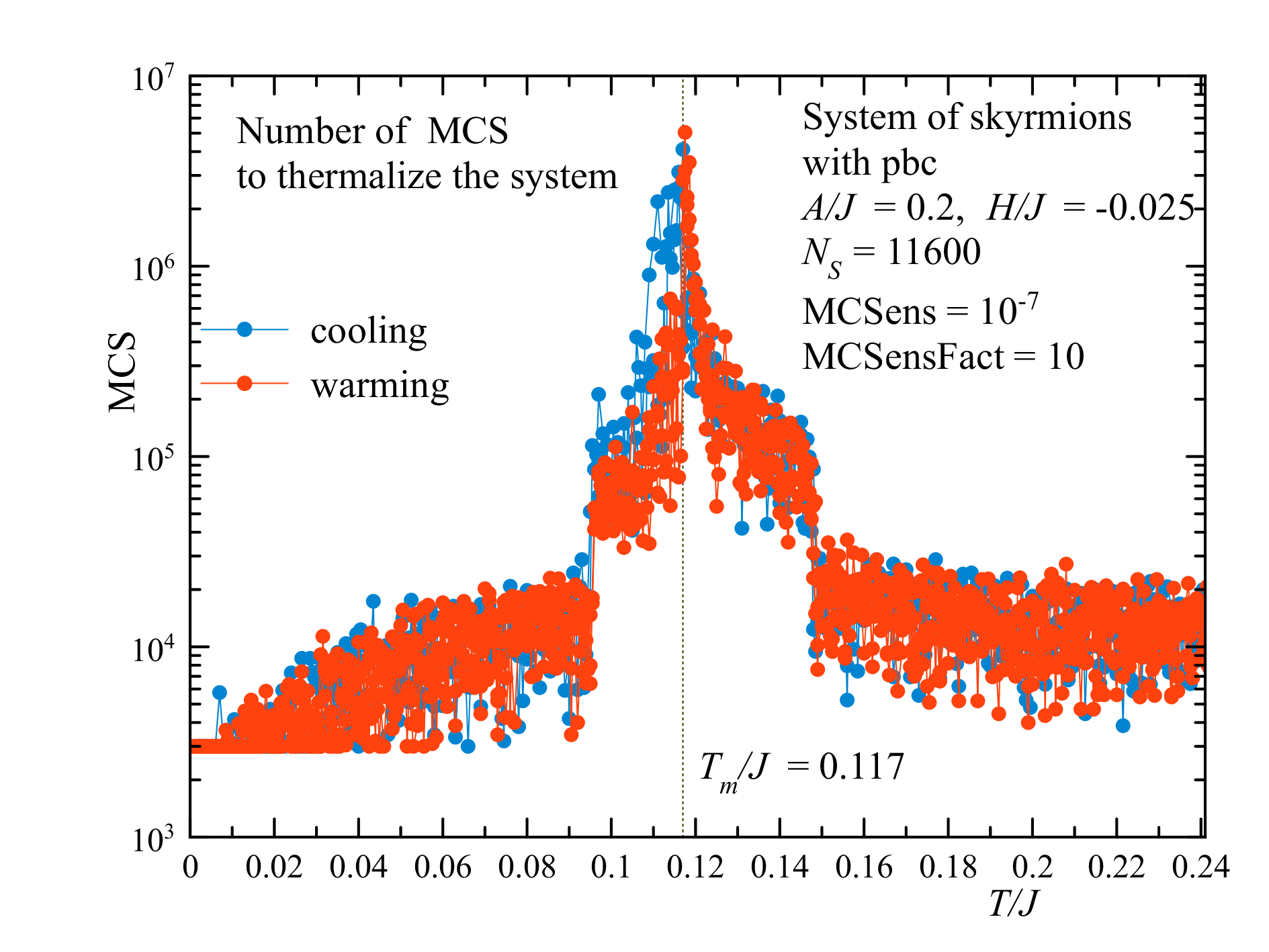}\caption{The number of MCS to thermalize the system strongly increases in the
region of melting/freezing.\label{Fig-nMCS_vs_T}}
\end{figure}

For computations in a broad region across the phase transition, it
is very important to employ a Monte Carlo routine with an automated
stopping criterion, as the number of MCS' necessary for equilibrating
the system and measuring the physical quantities changes by orders
of magnitude. Routines of this kind were used in some preceding work,
e.g., in Ref. \cite{GC-PRB2023}, but here we formulated the definitive
algorithm. First, one chooses the control quantity (CQ) which is nonzero
in both phases. One suitable quantity is the energy but here we use
$V_{6}$ defined by Eq. (\ref{V6_def}). The MC simulation is performed
in blocks of MCBlock Monte Carlo steps, typical values being $\mathrm{MCBlock}=50-100$.
After the completion of each block, the control quantity and other
physical quantities are computed. The next computational parameter
is $\mathrm{MCMass}<1$ which defines the volume of the computed data
used for data processing and finally for the measurement. We form
the lists of the most recent data of the length $L=\mathrm{Round}[\mathrm{MCMass}\times N_{\mathrm{val}}]$,
where $N_{\mathrm{val}}=\mathrm{MCS}/\mathrm{MCBlock}$ is the number
of computed values of the physical quantities. That is, we use the
fraction MCMass of all computed data, keeping the most recent results,
while older results are discarded. Both $N_{\mathrm{val}}$ and $L$
increase in the course of the simulation. Then, for the control quantity
we compute the moving averages over the lists of length $L$ using
the triangular averaging kernel $K_{L,i}$
\begin{equation}
\mathrm{QCmean}=\sum_{i=1}^{L}K_{L,i}\mathrm{QCList}_{i},\qquad\sum_{i=1}^{L}K_{L,i}=1,
\end{equation}
and form the list of the same length $L$ from these averages, CQmeanList.
The triangular averaging kernel yields a rather smooth list of data
which is used to quantify the evolution of the control quantity. As
$K_{L,i}$ has a maximum at $i=L/2$, the values from the middle of
the averaging interval $L$ make the largest contribution. 

We define the slope of the CQ per one MCS as
\begin{equation}
\mathrm{CQ}_{\mathrm{slope}}=\frac{\mathrm{First[CQmeanList]}-\mathrm{Last[CQmeanList]}}{L\times\mathrm{MCBlock}}\label{CQSlope_def}
\end{equation}
Then we define the slope parameter
\begin{equation}
\mathrm{SlopePar}=\mathrm{CQ}_{\mathrm{slope}}/\left(\mathrm{CQ}_{0}\times\mathrm{MCSens}\right).\label{SlopePar_def}
\end{equation}
Here $\mathrm{\mathrm{CQ}_{0}}$ is the typical value of the control
quantity, $\mathrm{CQ}_{0}=1$ for $V_{6}$, and MCSens is the sensitivity
of our Monte Carlo simulation, a very small number. We form the list
of the most recent values of $\left|\mathrm{SlopePar}\right|$ of
the same length $L$. The simulation is stopped when the mean value
of this list becomes smaller than one -- this is the most important
new feature of the method. 

\begin{figure}
\begin{centering}
\includegraphics[width=8cm]{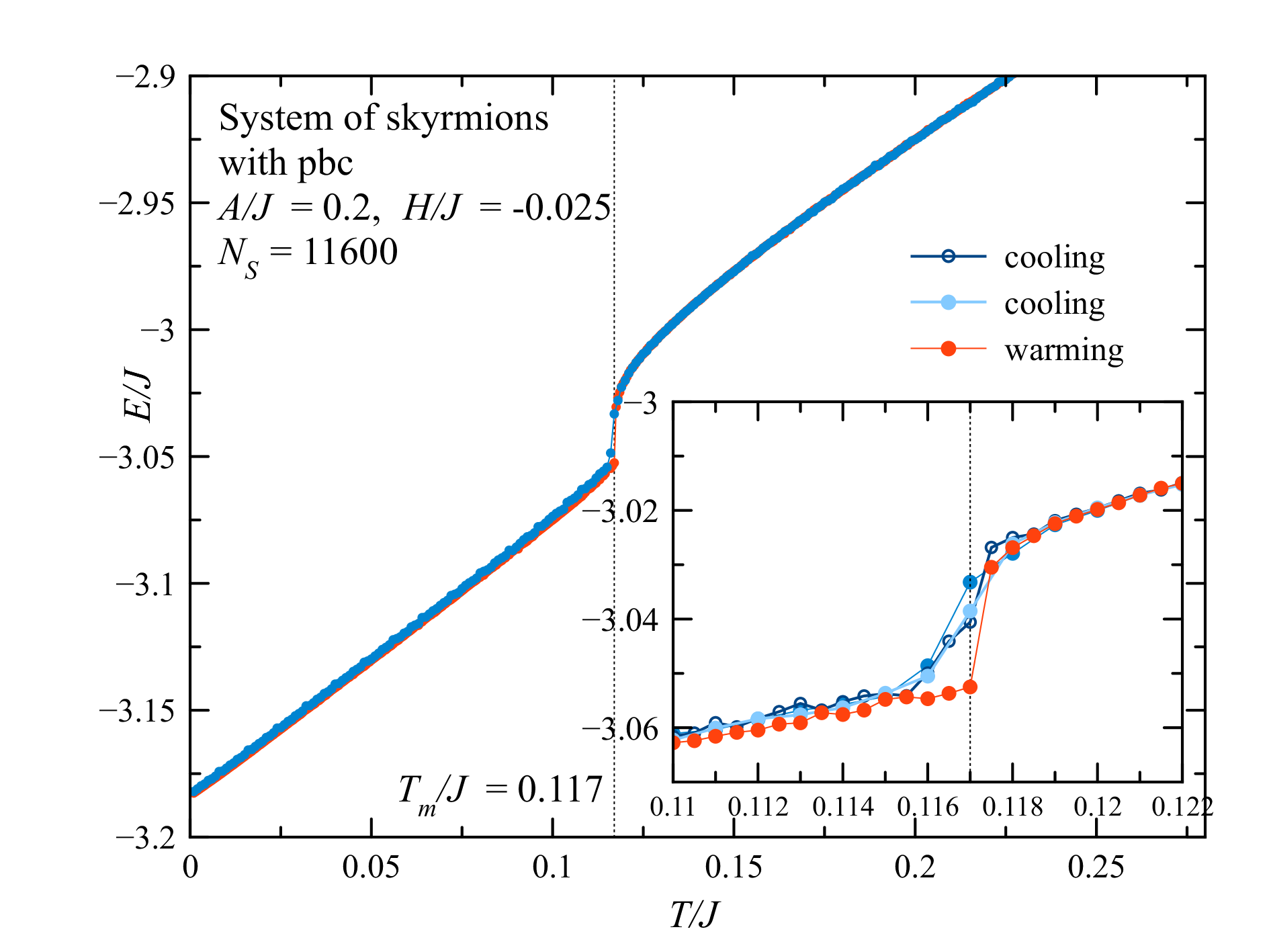}
\par\end{centering}
\begin{centering}
\includegraphics[width=8cm]{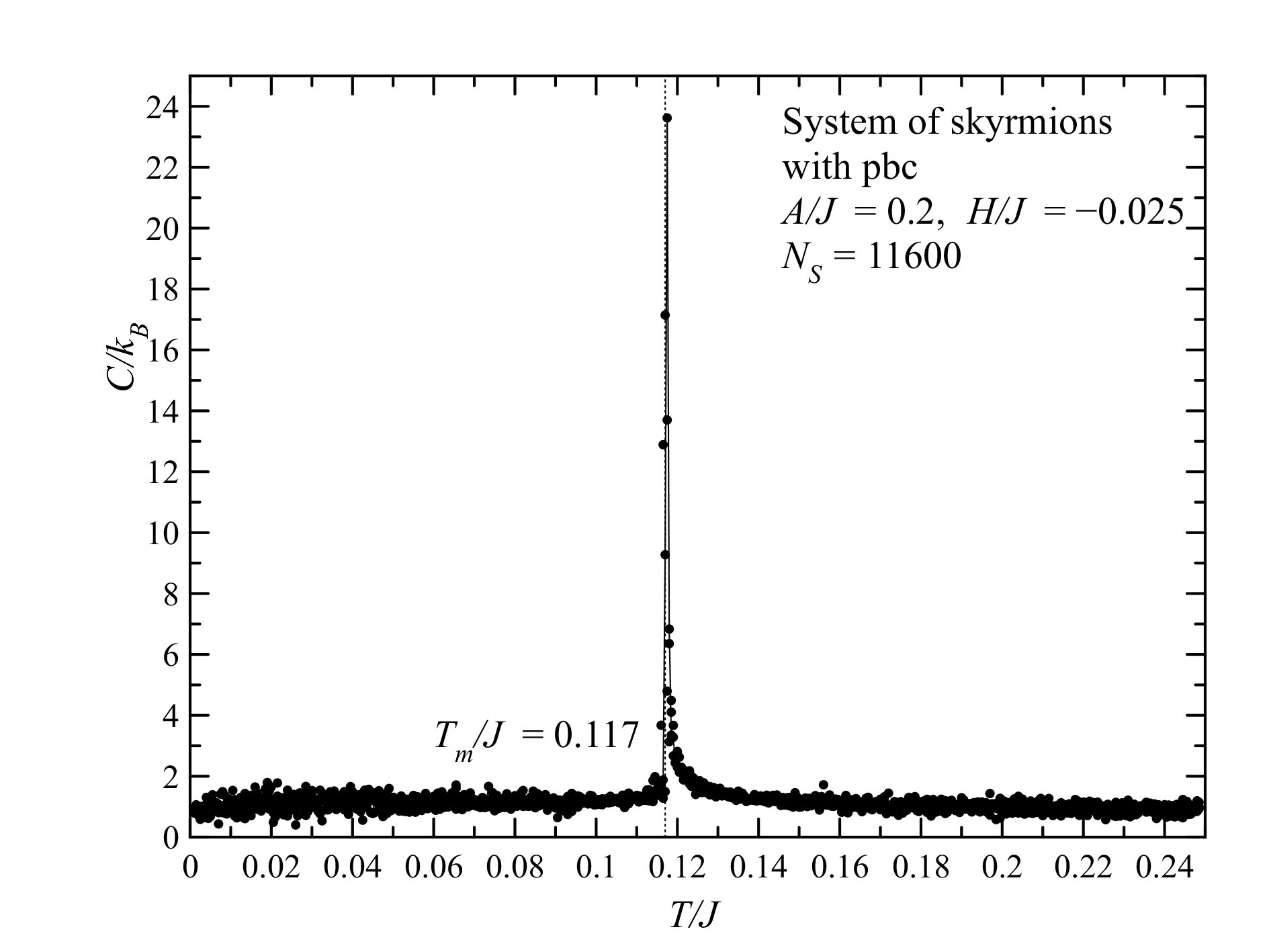}
\par\end{centering}
\caption{The temperature dependences of the energy per skyrmion (Upper panel.
Inset shows the magnification of the transition region with a negligible
hysteresis) and the heat capacity per skyrmion obtained from the energy
fluctuations, Eq. (\ref{C_def}) (Lower panel).}

\label{Fig-E_vs_T}
\end{figure}
\begin{figure}
\begin{centering}
\includegraphics[width=8cm]{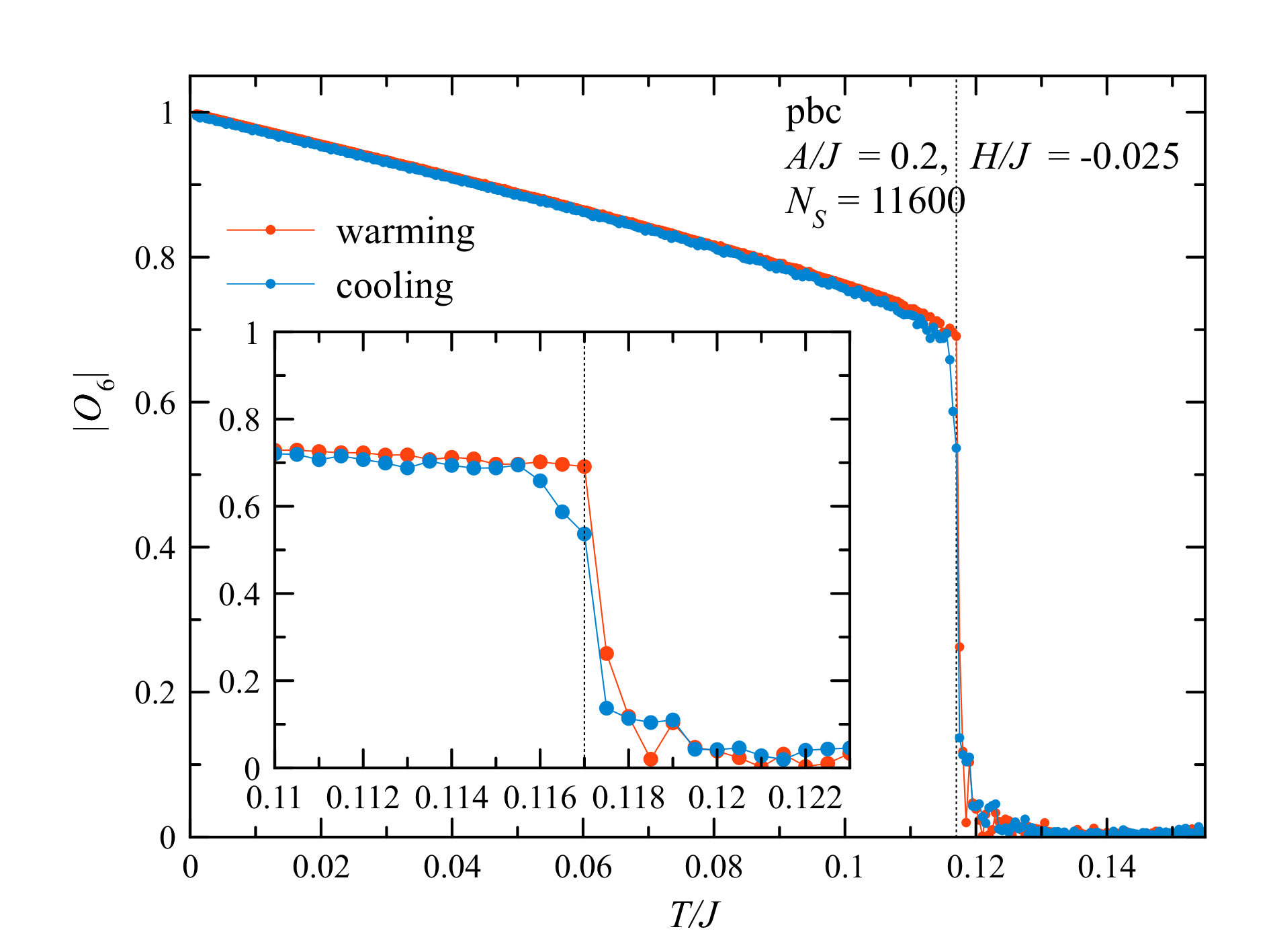}
\par\end{centering}
\begin{centering}
\includegraphics[width=8cm]{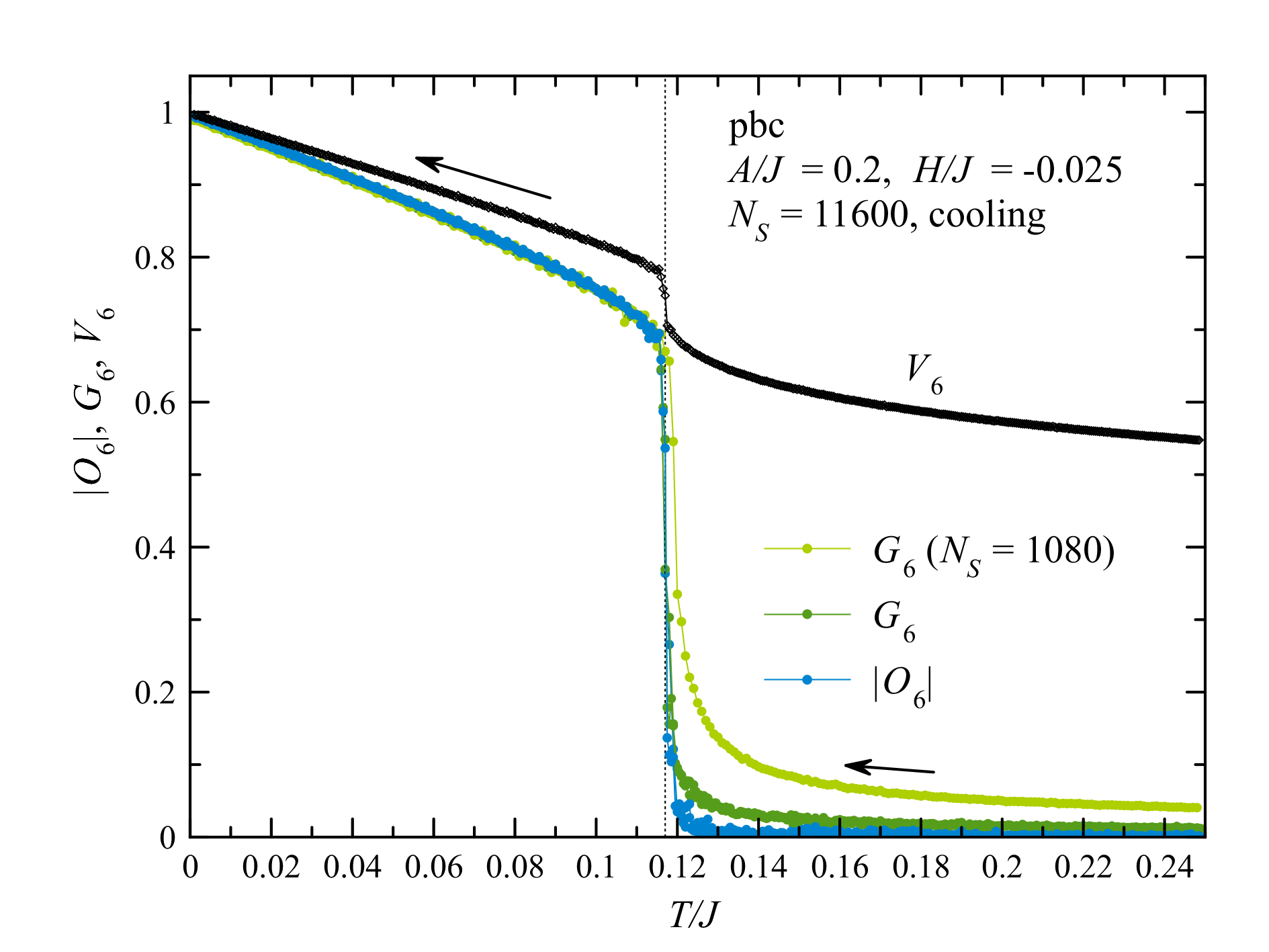}
\par\end{centering}
\caption{Orientational order parameter $O_{6}$ vs temperature (Upper panel.
Inset shows a jump with a negligible hysteresis) and the temperature
dependence of the orientational quantities on cooling (Lower panel).}

\label{Fig-O6_vs_T}
\end{figure}

One more improvement makes an accent on the critical region that is
defined as the region where $V_{6,\mathrm{floor}}<V_{6}<V_{6,\mathrm{ceiling}}$.
We used $V_{6,\mathrm{floor}}^{2}=0.38$ and $V_{6,\mathrm{ceiling}}^{2}=0.69$.
These bracket the region in which melting/freezing takes place, $V_{6}^{2}\simeq0.5$.
Whenever the computed value of $V_{6}$ falls into this interval,
SlopePar is multiplied by $\mathrm{MCSensFact}=10$. As a result,
the stopping criterion becomes more stringent in the critical region
than otherwise. The method always converges, even in the case of slow
large-amplitude stationary fluctuations because of the increasing
$L$ in the definition of $\mathrm{CQ}_{\mathrm{slope}}$, Eq. (\ref{CQSlope_def}).
After stopping, the latest of the moving averages of the control quantity,
Last{[}CQmeanList{]}, is output. For the other quantities, the mean
values over the lists of $L$ most recent values are output. The smaller
the parameter MCSens, the longer simulation is done and the more extensive
is the averaging of the computed quantities. Fig. \ref{Fig-nMCS_vs_T}
illustrates the number of MCS needed to achieve thermalization for
the system of 11600 skyrmions with $\mathrm{MCSens}=10^{-7}$. The
number of MCS increases in the critical region and peaks at the melting/freezing
point $T/J=0.117$. Since the main computing time is spent near melting/freezing
anyway, one can make computations in a broad range of $T$ with the
same $T$-interval. One does not need to rarify the points away from
the phase transition region because these points cost a relatively
short computing time.

The Monte Carlo algorithm presented above is superior to the version
used in Ref. \cite{GC-PRB2023}. Whereas the number of MCS done in
Ref. \cite{GC-PRB2023} does not exceed one million, here it goes
up to several millions that provides a better thermalization and averaging
of stationary fluctuations near the transition. The main focus here
is on the system with $\sim10^{4}$ skyrmions which is sufficiently
large to study thermodynamics. Here in computing temperature dependences
of the quantities, we used a finer $T$-grid than in Ref. \cite{GC-PRB2023}.

Computations were performed with Wolfram Mathematica using compilation
of the core routines such as Monte Carlo and those computing the physical
quantities. Dependences on $T$ and $H$ were computed in parallelized
cycles using the number $N_{\mathrm{cores}}$ of the available processor
cores and the so-called \textit{relay }method (see Fig. \ref{Fig-Relay_computation})
in which the next $T$-point within the same thread uses the previously
obtained state as the initial condition. With the interval between
the neighboring $T$-points being $\delta T$, the interval between
those within the same thread is $\Delta T=N_{\mathrm{cores}}\delta T$
that is sufficiently small if $\delta T$ is small and $N_{\mathrm{cores}}$
is not too large. We mainly used 8 cores for these computations. The
initial conditions were perfect lattice or random placement of skyrmions.

\section{Thermodynamics of melting and freezing}

\label{Sec_Thermodynamics}

\begin{figure}
\begin{centering}
\includegraphics[width=8cm]{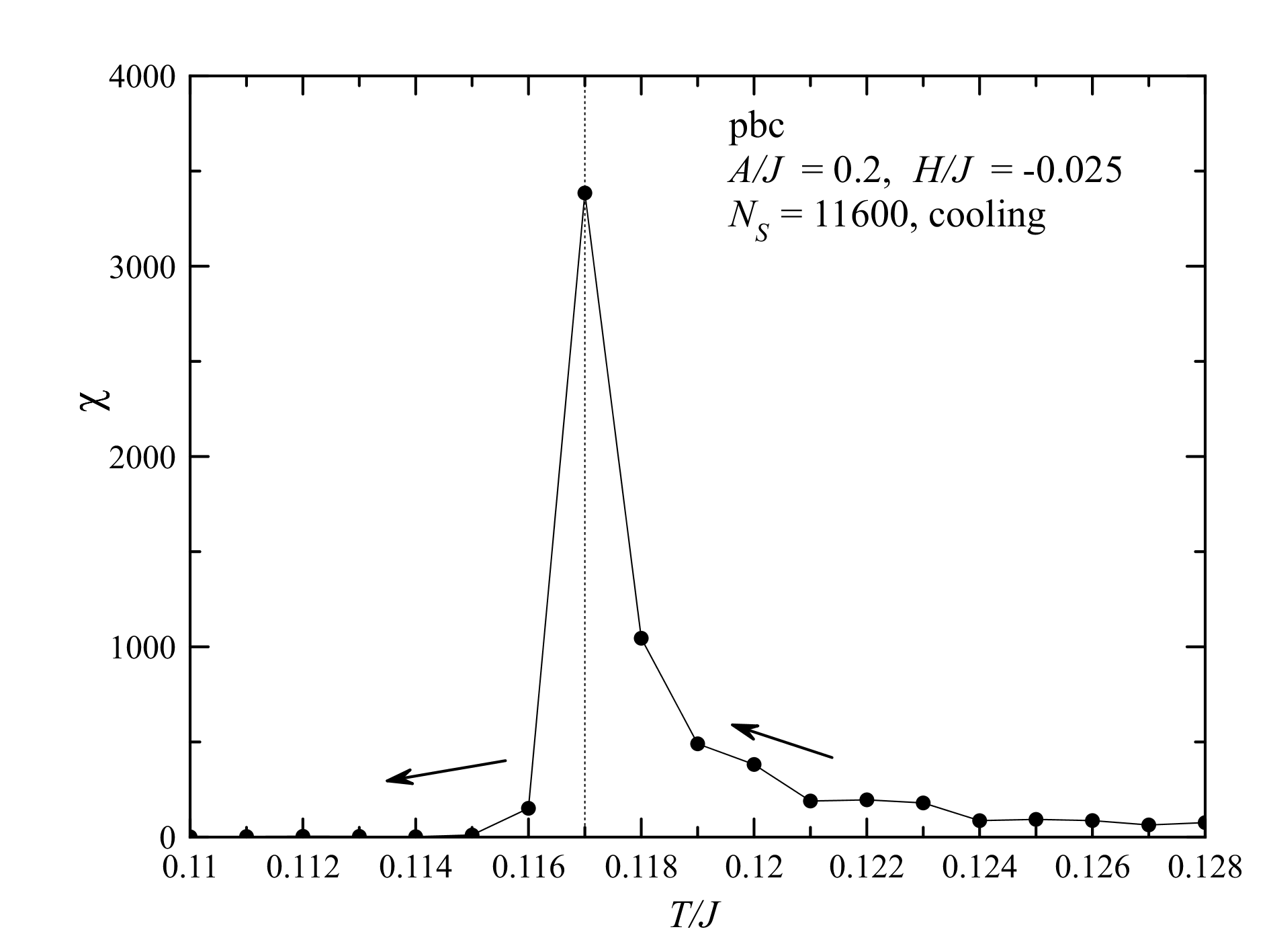}
\par\end{centering}
\begin{centering}
\includegraphics[width=8cm]{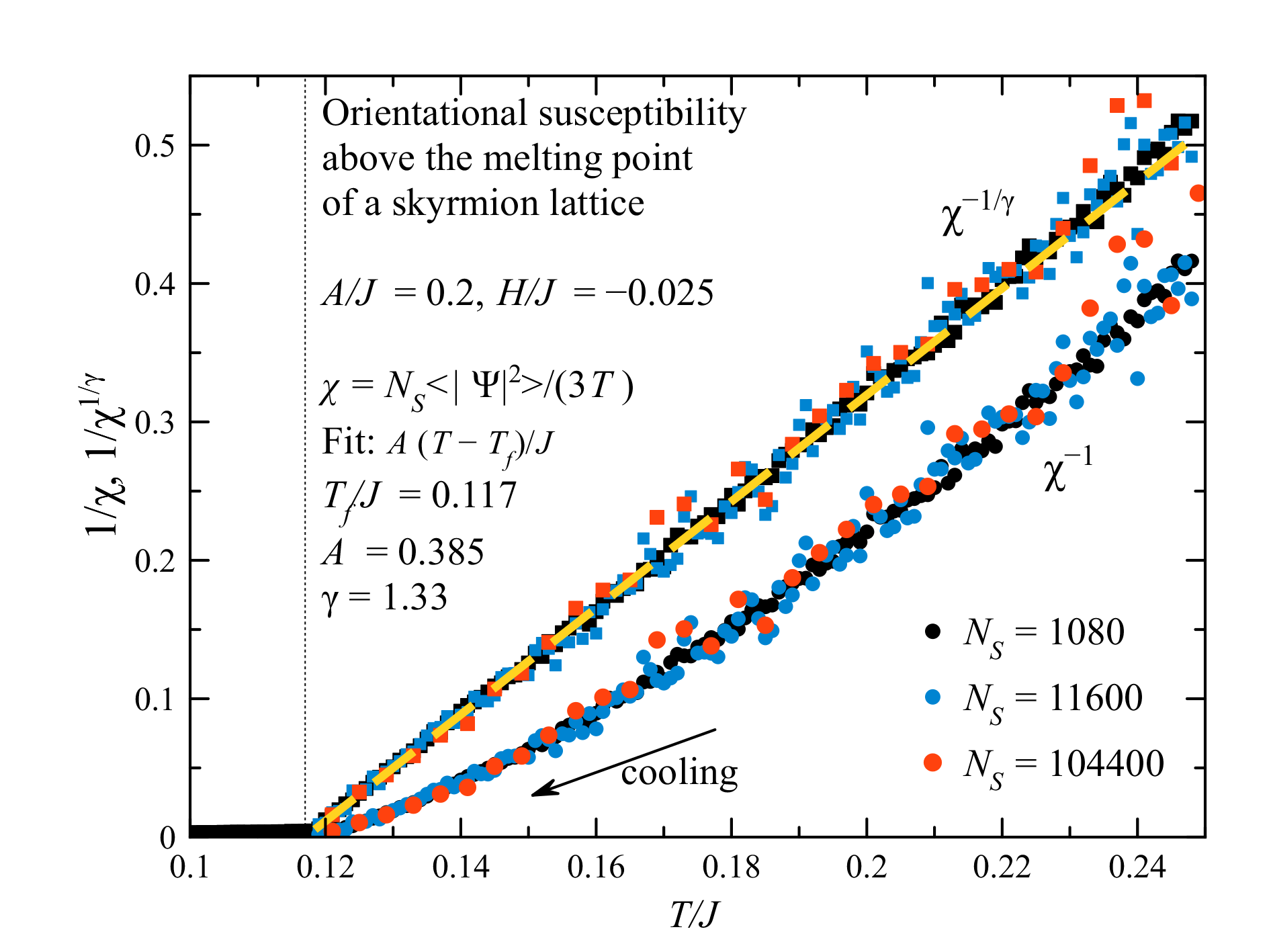}
\par\end{centering}
\caption{Upper panel: Susceptibility associated with the orientational order
peaks at the melting/freezing transition. Lower panel: Critical dependence
of the inverse susceptibility above freezing.}

\label{Fig-chi_vs_T}
\end{figure}
\begin{figure}
\begin{centering}
\includegraphics[width=8cm]{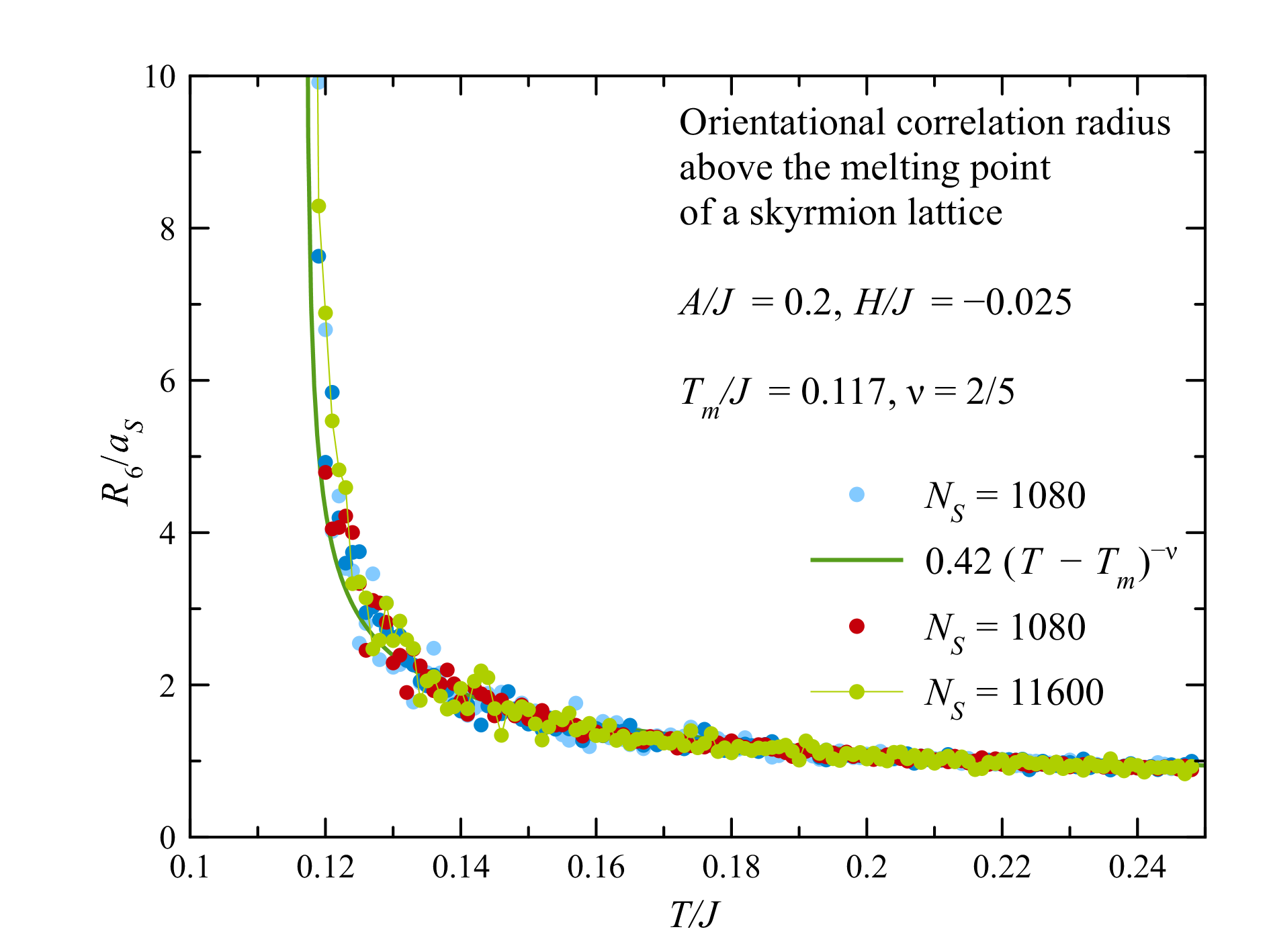}
\par\end{centering}
\begin{centering}
\includegraphics[width=8cm]{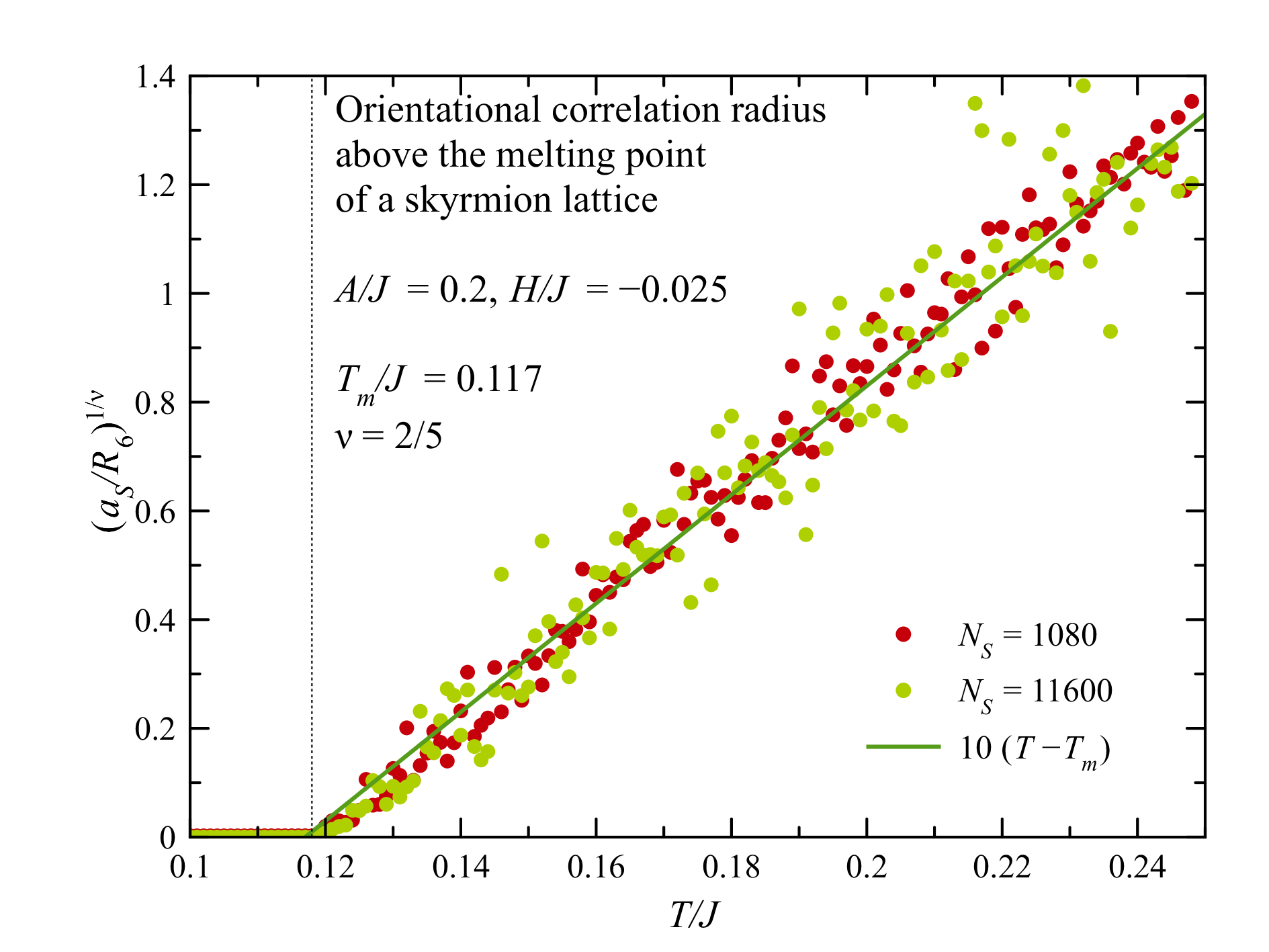}
\par\end{centering}
\caption{Orientational correlation length $R_{6}$ computed using Eq. (\ref{R6}). }

\label{Fig-R6_vs_T}
\end{figure}

We have found that in our case SkL melting/freezing occurs according
to the grain scenario \cite{Chui,Saito} via only one melting/freezing
transition between solid and liquid phases. In Ref. \cite{GC-PRB2023}
we concluded that this phase transition is first order because the
orientational order parameter $O_{6}$ jumped to zero at the melting
temperature. However, as we show here, this transition possesses features
of second-order transitions such as divergent heat capacity and susceptibility.
These are more reliable in the determination of the order of the transition
than a jump of the order parameter because in simulations it is difficult
to distinguish a jump from a very steep dependence.

Melting/freezing is a very slow process including slow relaxation
and slow large-amplitude stationary fluctuations. Large areas of the
system are melting and freezing again. Nevertheless, for the system
of $10^{4}$ skyrmions thermodynamic averages at equilibrium can be
obtained by the Metropolis Monte Carlo within $10^{6}-10^{7}$ Monte
Carlo steps (MCS), see Fig. \ref{Fig-nMCS_vs_T}. With a full thermalization,
there is practically no hysteresis and the melting temperature $T_{m}$
is practically equal to the freezing temperature $T_{f}$. Hysteresis
is observed in the cases of incomplete equilibration, for instance,
if the stopping criterion for thermalization is too liberal or if
the temperature changes at a too high rate.

Larger systems, such as $10^{5}$ skyrmions, tend to freeze into a
polycrystal state which then evolves very slowly toward the monocrystal
state, see Sec. \ref{Sec_Nonequilbrium}. This process for large systems
is even slower than melting/freezing. 

\begin{figure}
\begin{centering}
\includegraphics[width=8cm]{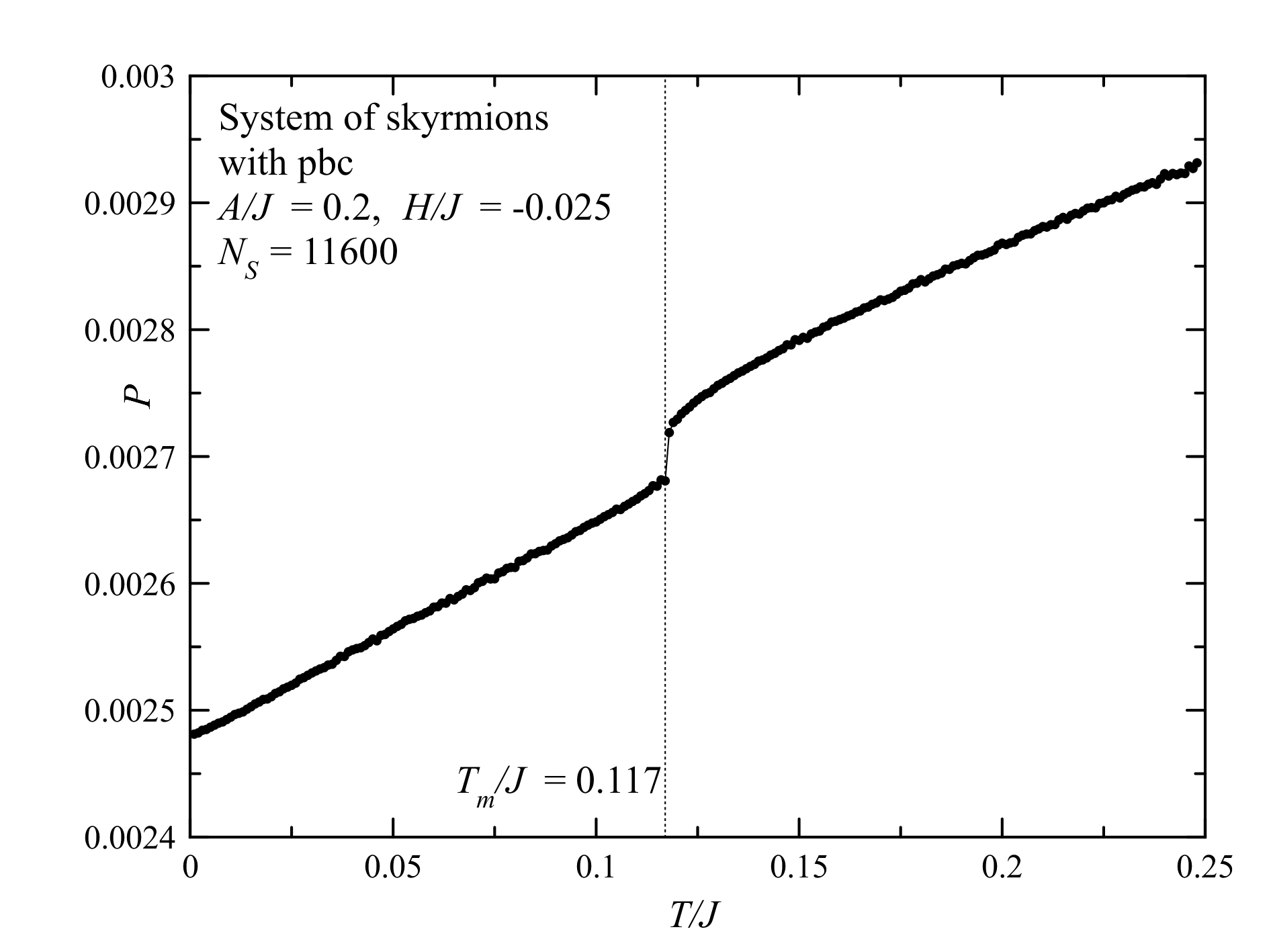}
\par\end{centering}
\caption{Temperature dependence of the pressure in the skyrmion matter.}

\label{Fig-P_vs_T}
\end{figure}
\begin{figure}
\begin{centering}
\includegraphics[width=8cm]{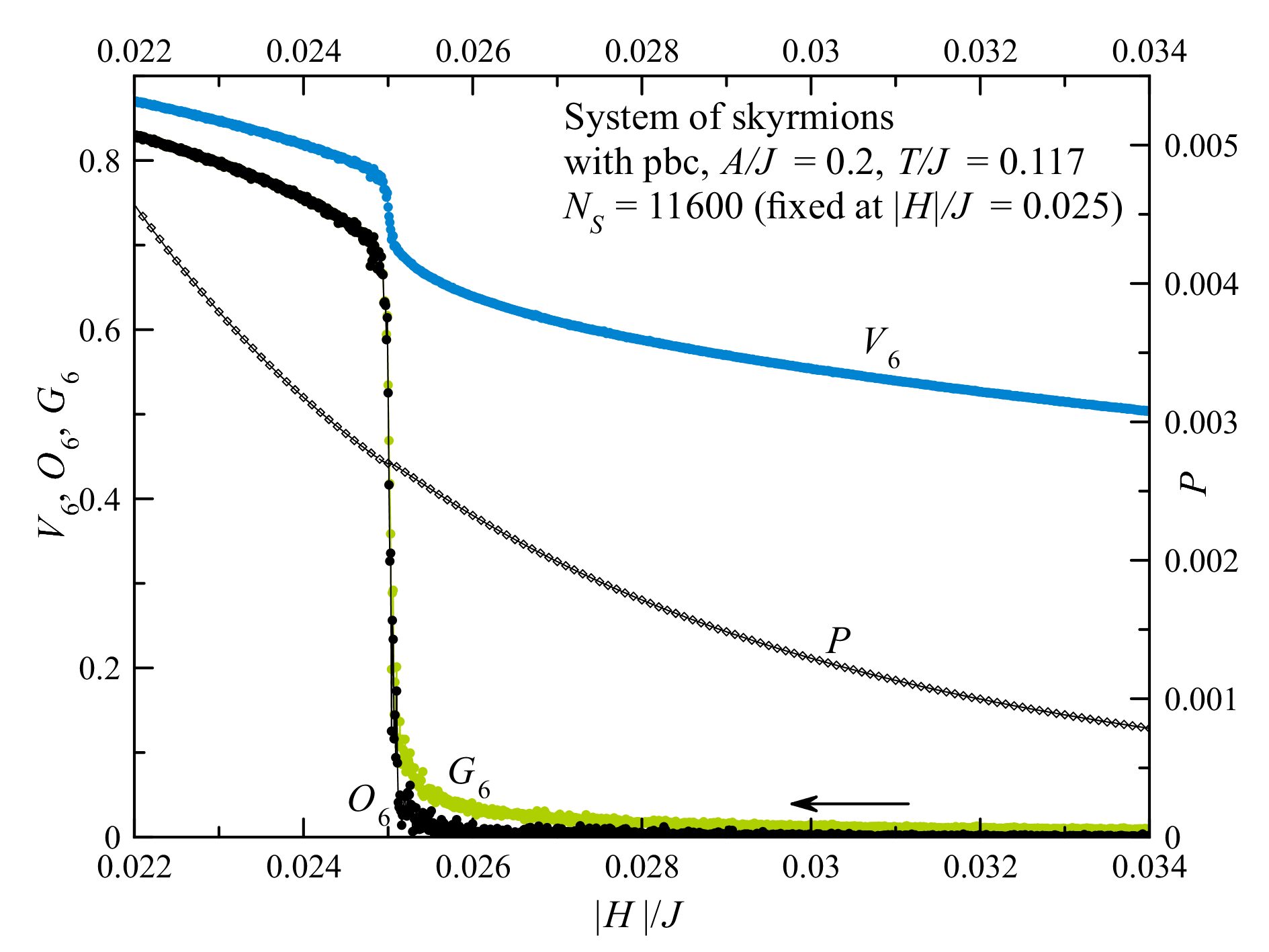}
\par\end{centering}
\begin{centering}
\includegraphics[width=8cm]{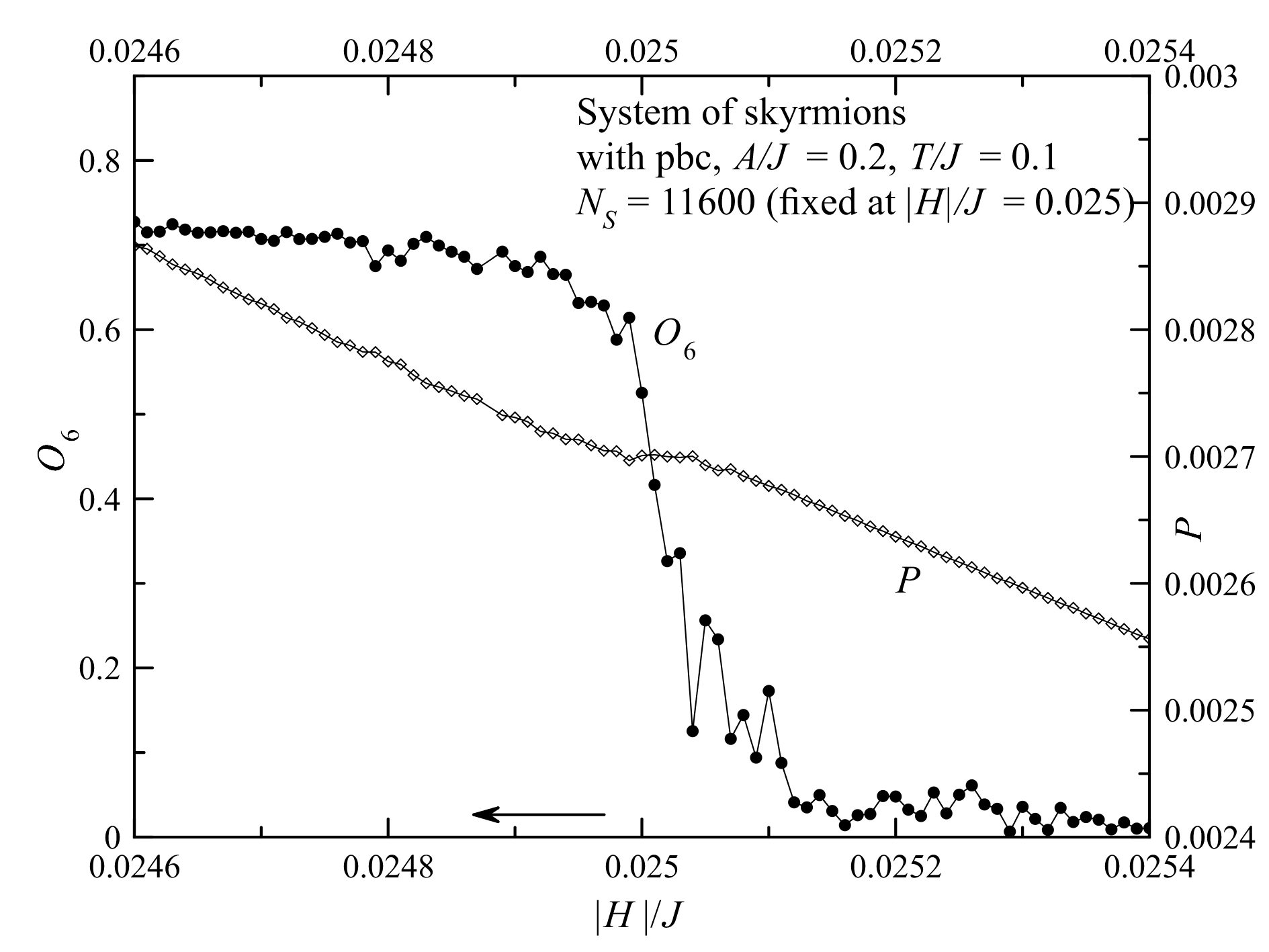}
\par\end{centering}
\caption{Freezing of the skyrmion lattice on decreasing the strength of the
magnetic field\textbf{ }is a continuous transition with a small Mayer-Wood
loop for the pressure.}

\label{Fig-P_vs_H}
\end{figure}

The energy per particle $E$ shows a singularity just above the transition,
see Fig. \ref{Fig-E_vs_T}. The heat capacity $C$ per particle obtained
from the energy fluctuations, Eq. (\ref{C_def}), and shown in the
lower panel of Fig. \ref{Fig-E_vs_T} peaks in this region, although
the numerical data does not allow to reliably infer its critical behavior.
The singularity of $C$ is not a robust power law and resembles a
logarithmic divergence. In the low-temperature region the analytical
result is $C=k_{B}$ that is reproduced in the simulation.

The orientational order parameter $O_{6}$ exhibits a linear dependence
at low temperatures (that is typical for classical systems) and a
jump or an unresolved very steep dependence at melting/freezing, see
Fig. \ref{Fig-O6_vs_T}. The average hexagonality value $V_{6}$ decreases
with the temperature and shows a steep descent just above the melting/freezing
point (lower panel of Fig. \ref{Fig-O6_vs_T}). The orientational
parameter $G_{6}$ is smooth above melting and decreases with the
temperature and the system size. In the skyrmion-liquid phase, $G_{6}\gg O_{6}$,
while in the solid phase $G_{6}\cong O_{6}$, thus the susceptibility
has a peak at melting/freezing, as is shown in the upper panel of
Fig. \ref{Fig-chi_vs_T}. To obtain this plot, we used $\mathrm{MCSence}=10^{-8}$
with $\mathrm{MCSensFact}=10$. The inverse susceptibility computed
for the systems of $10^{3}$, $10^{4}$, and $10^{5}$ skyrmions and
shown in the lower panel of the same figure has an apparent critical
behavior with the critical index $\gamma\simeq1.33$.

Fig. \ref{Fig-R6_vs_T} shows the temperature dependence of the orientational
correlation length above melting, obtained from $G_{6}$, Eq. (\ref{G6_def}).
The exponential Ansatz for the orientational correlation function,
see Eq. (\ref{R6}), yields a critical divergence of the correlation
length $R_{6}/a_{S}\simeq0.42\left[(T-T_{m})/J\right]^{-\nu}$ with
$\nu=2/5$.

\begin{figure}
\begin{centering}
\includegraphics[width=8cm]{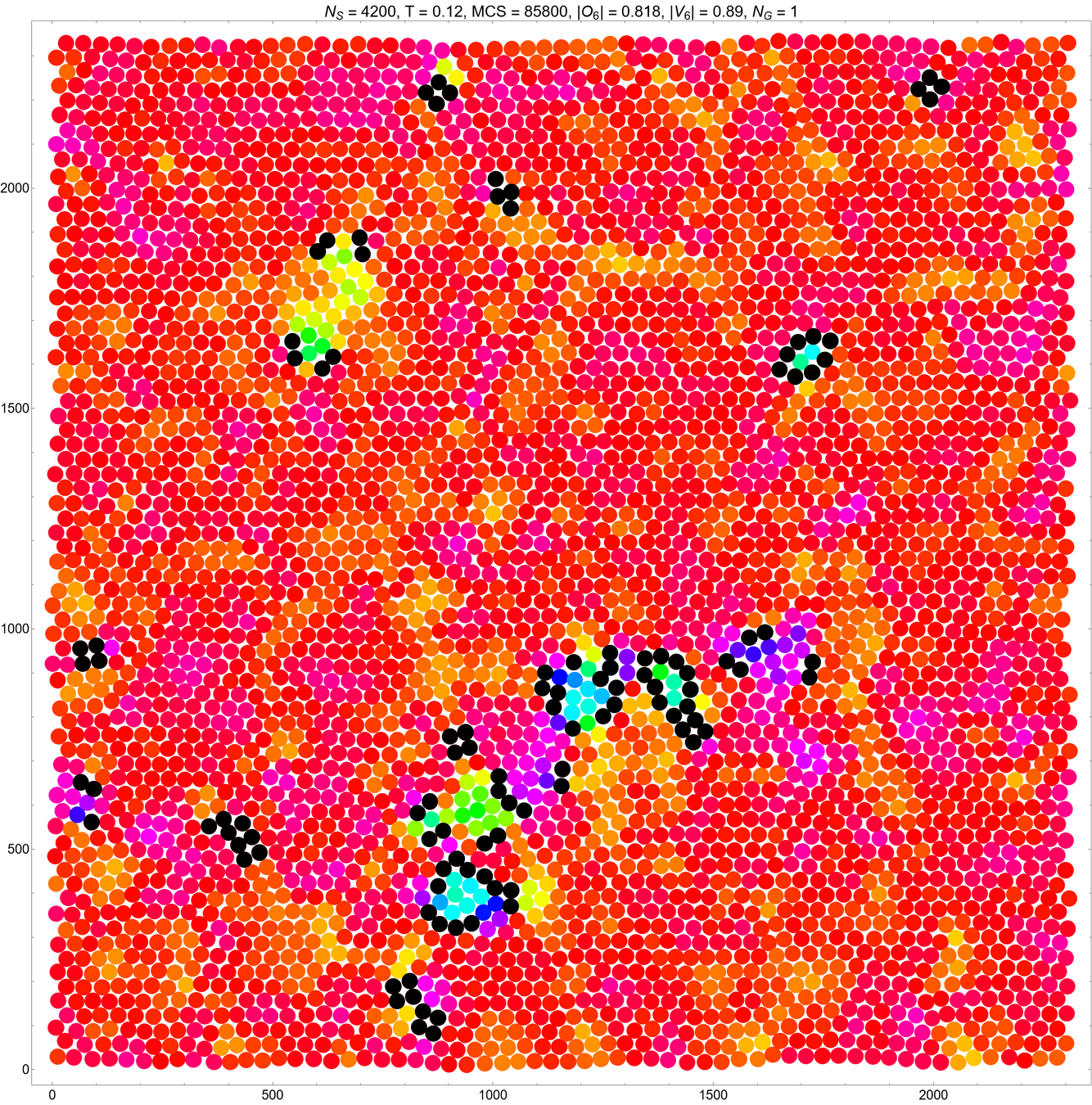}
\par\end{centering}
\begin{centering}
\includegraphics[width=8cm]{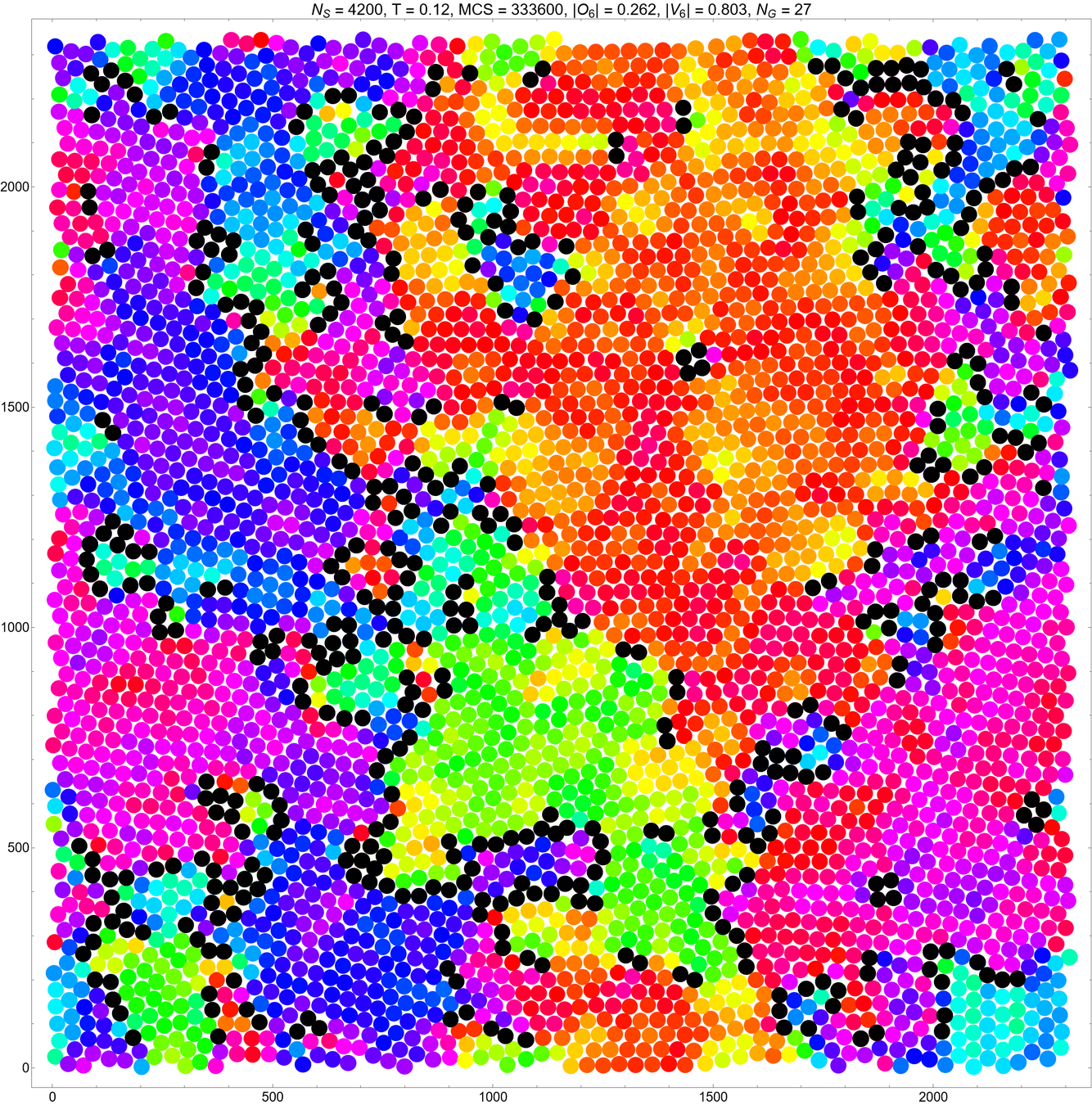}
\par\end{centering}
\caption{Early and advanced stages of melting of the SkL slightly above the
melting temperature. The process goes via formation of grains with
a different orientations of hexagons (color coded). Grains are separated
by lines of dislocations shown in black.}

\label{Fig-melting}
\end{figure}

The system of skyrmions is a system with a pure repulsion, thus it
is always under pressure that keeps the system together. One can calculate
pressure as the force acting perpendicularly across any line in the
system per unit length of this line. For a perfect lattice at $T=0$
taking into account only the six nearest neighbors one obtains
\begin{equation}
P=\frac{\sqrt{3}f}{a_{S}},\qquad f=\left.\frac{dU(r)}{dr}\right|_{r=a_{S}}=\frac{F}{\delta_{H}}\exp\left(-\frac{a_{S}}{\delta_{H}}\right).\label{P_def}
\end{equation}
where $f$ is the magnitude of the repulsion force between the two
skyrmions. The temperature dependence of the pressure shown in Fig.
\ref{Fig-P_vs_T} is similar to that of the energy, Fig. \ref{Fig-E_vs_T}.
In particular, melting causes a steep increase of the pressure. 

Usually pressure is computed for different concentrations of particles.
For a system of skyrmions, it is more convenient to control pressure
by the applied magnetic field as the repulsion energy depends on $H$
via the magnetic length $\delta_{H}=\sqrt{J/|H|}$ and the exponential
prefactor $F$. Increasing the field strength reduces the interaction
and causes melting of the skyrmion lattice, as shown in Fig. \ref{Fig-P_vs_H}.
The dependence on $H$ is less steep than the temperature dependence,
and one can see that the phase transition on $H$ is clearly continuous.

Increasing the strength of the magnetic field reduces the pressure
in the system. However, melting of the skyrmion lattice increases
pressure that can create a Mayer-Wood loop. This loop is difficult
to see, though, as it is masked by a strong regular part of the dependence
$P(H)$ that is much more pronounced than the $P(T)$ dependence.
Still, one can see a small Mayer-Wood loop in the lower panel of Fig.
\ref{Fig-P_vs_H}.

\section{Spontaneous emergence of lattice defects and formation of grains}

\label{Sec_Defects}

\begin{figure}
\begin{centering}
\includegraphics[width=8cm]{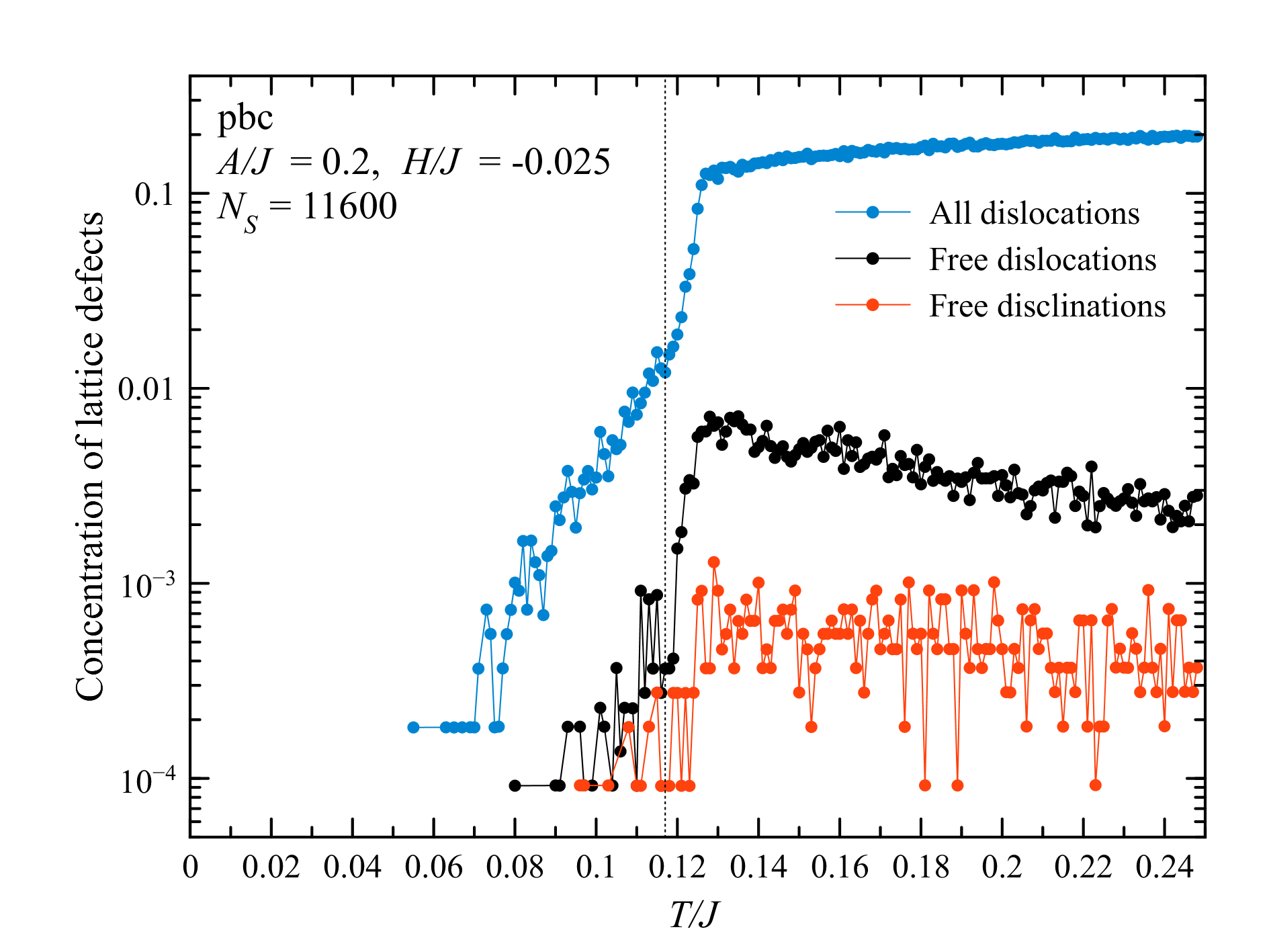}
\par\end{centering}
\caption{Temperature dependence of the number of lattice defects across the
melting transition.}

\label{Fig-cDisl_vs_T}
\end{figure}
\begin{figure}
\begin{centering}
\includegraphics[width=8cm]{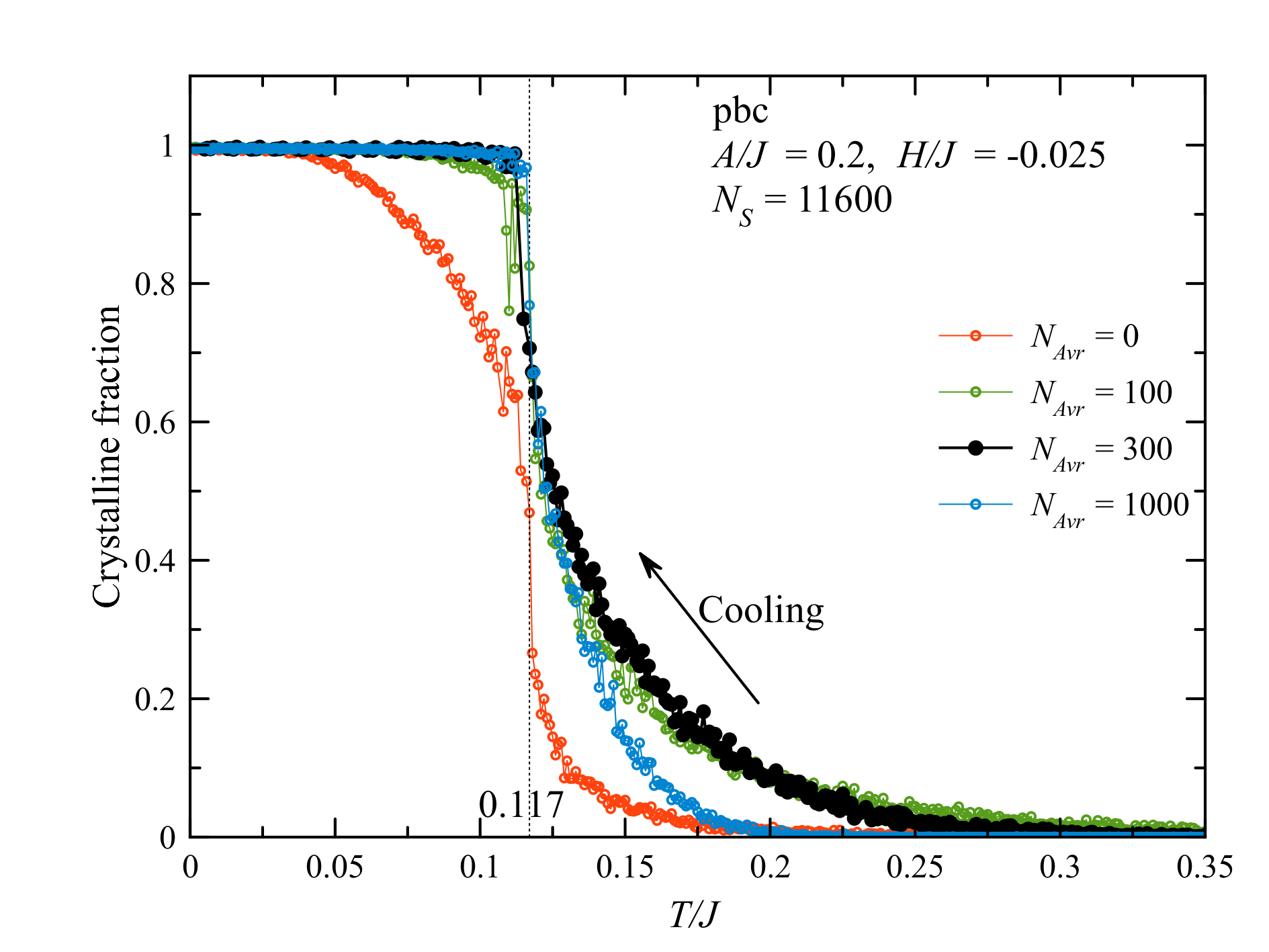}
\par\end{centering}
\caption{Crystal fraction vs temperature for different lengths of positional
averaging.}

\label{Fig-CrystFrac}
\end{figure}
\begin{figure}
\begin{centering}
\includegraphics[width=8cm]{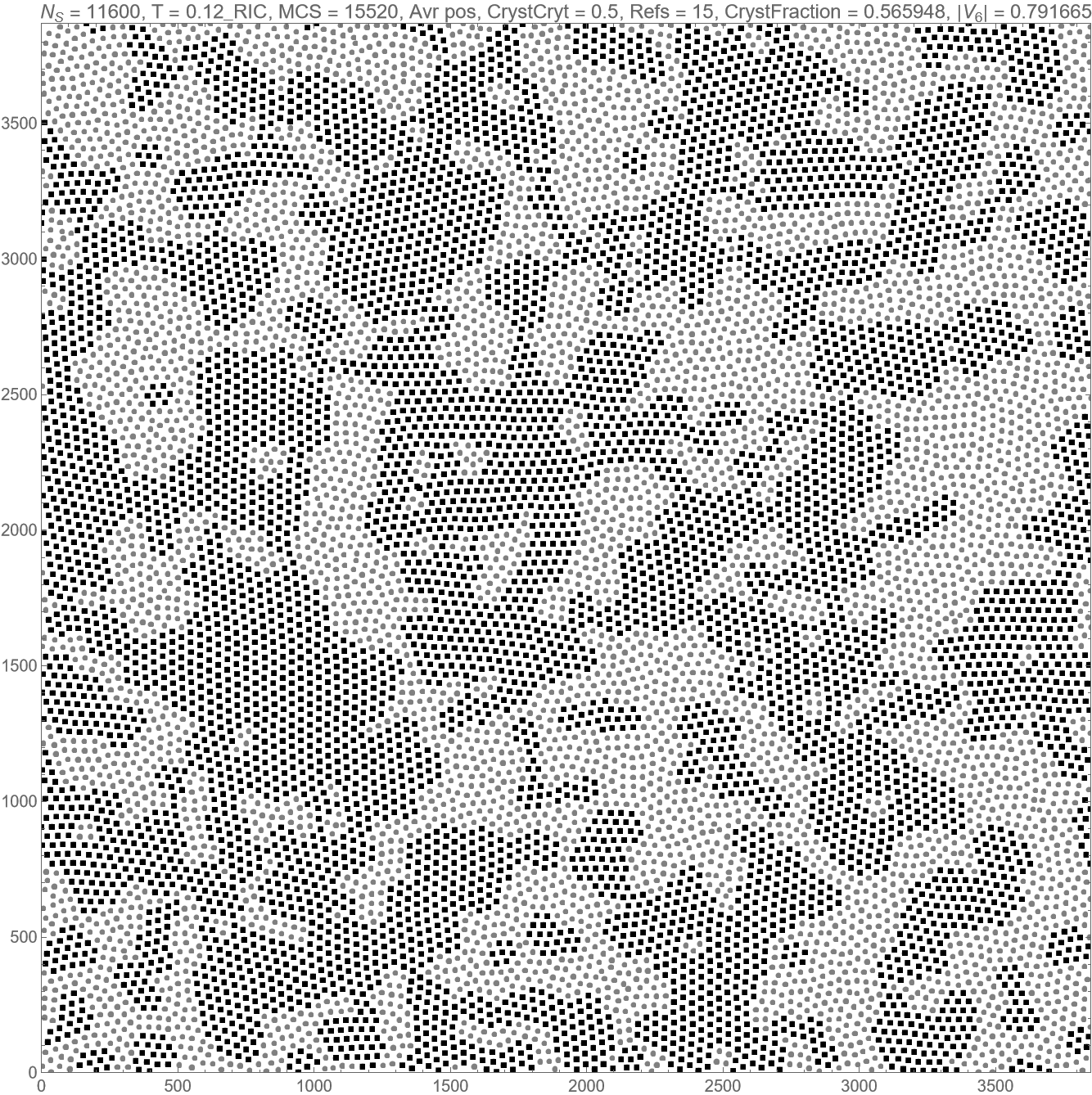}
\par\end{centering}
\begin{centering}
\includegraphics[width=8cm]{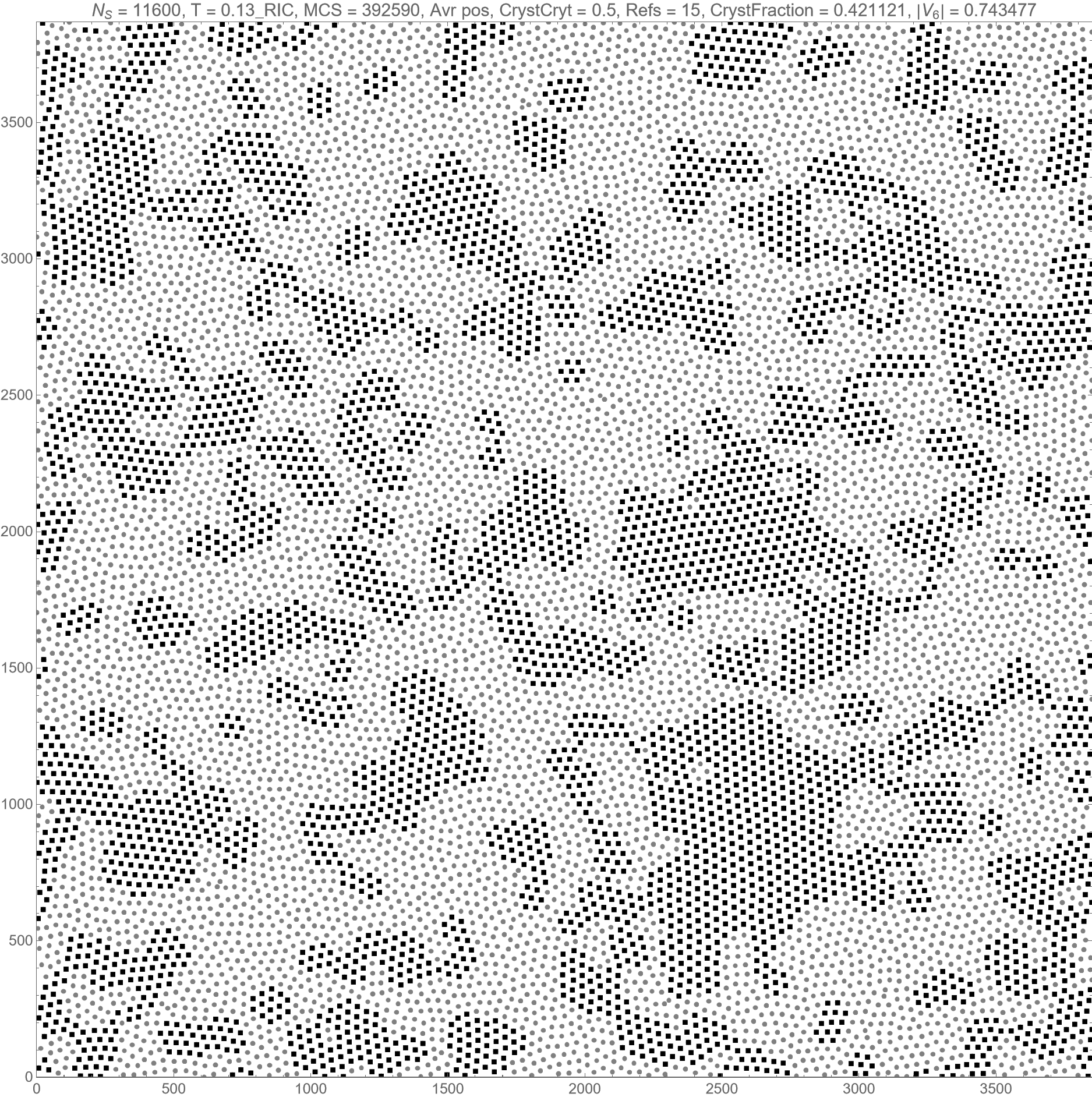}
\par\end{centering}
\caption{Skyrmion matter after positional averaging at $T/J=0.12$ and 0.13.
Skyrmions belonging to the solid part are shown by black points and
those belonging to the liquid part are shown by gray points.}

\label{Fig-Averaged_lattice}
\end{figure}

The process of melting of a skyrmion lattice goes via spontaneous
formation of grains with different orientation of hexagons, as suggested
by Chui (see Fig. \ref{Fig-melting}). Grains are separated from each
other by the lines where the lattice is broken and dislocations are
clustered. Clustered dislocations can be identified as skyrmions having
the number of Delaunay nearest neighbors different from six. Near
the melting/freezing temperature, in the course of the Monte Carlo
simulation, grains appear and disappear, grain boundaries move back
and forth. This process is very long. Above the melting temperature,
the tendency of increasing the number of differently oriented grains
dominates. Below this temperature, to the contrary, grains tend to
coalesce into bigger ones, until there is a uniform orientation of
hexagons in the system. We do not observe the so-called ``unbinding''
of dislocations or disclinations -- with increasing the temperature
lattice defects emerge bound in clusters.

\begin{figure}
\begin{centering}
\includegraphics[width=8cm]{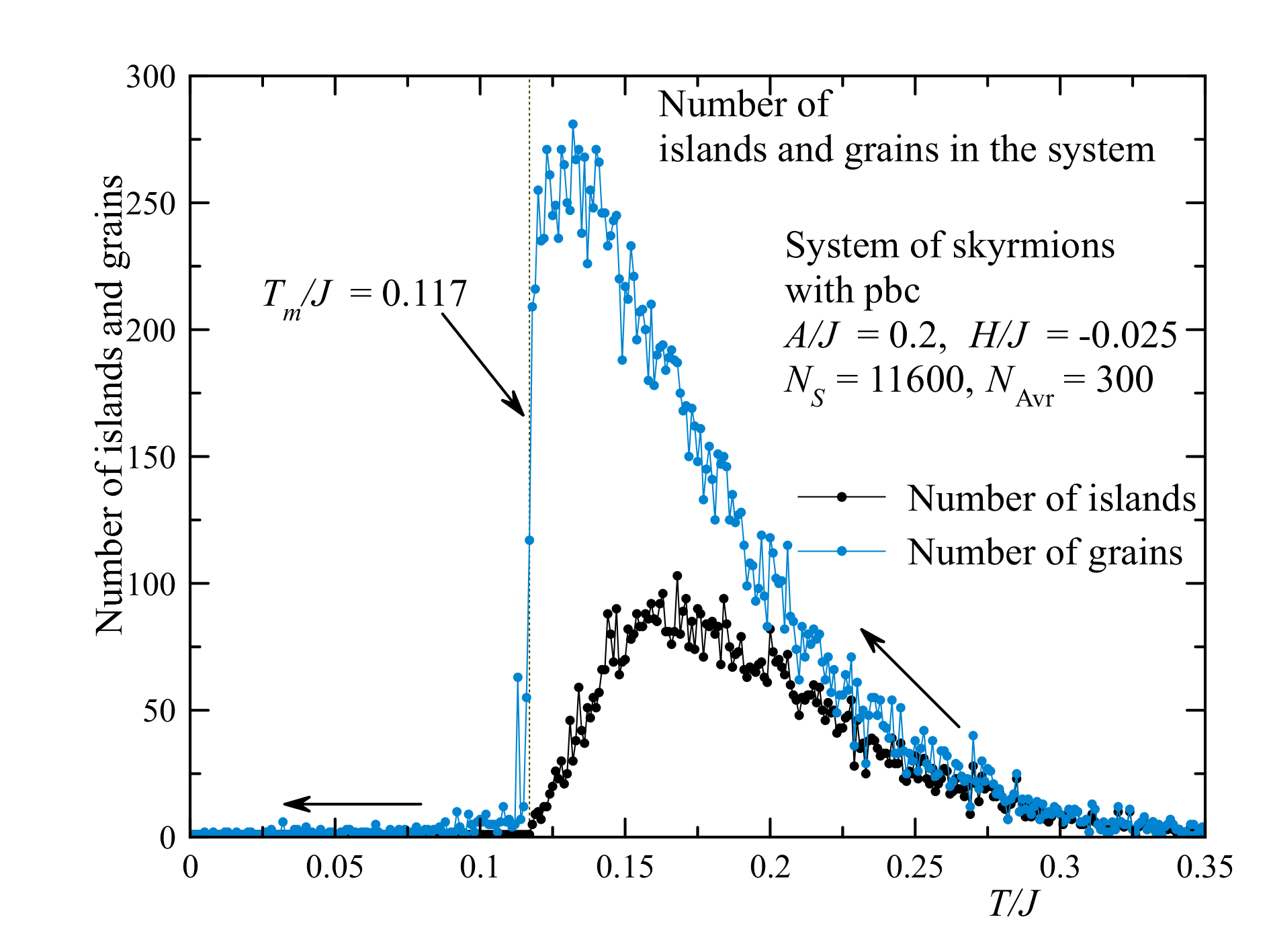}
\par\end{centering}
\begin{centering}
\includegraphics[width=8cm]{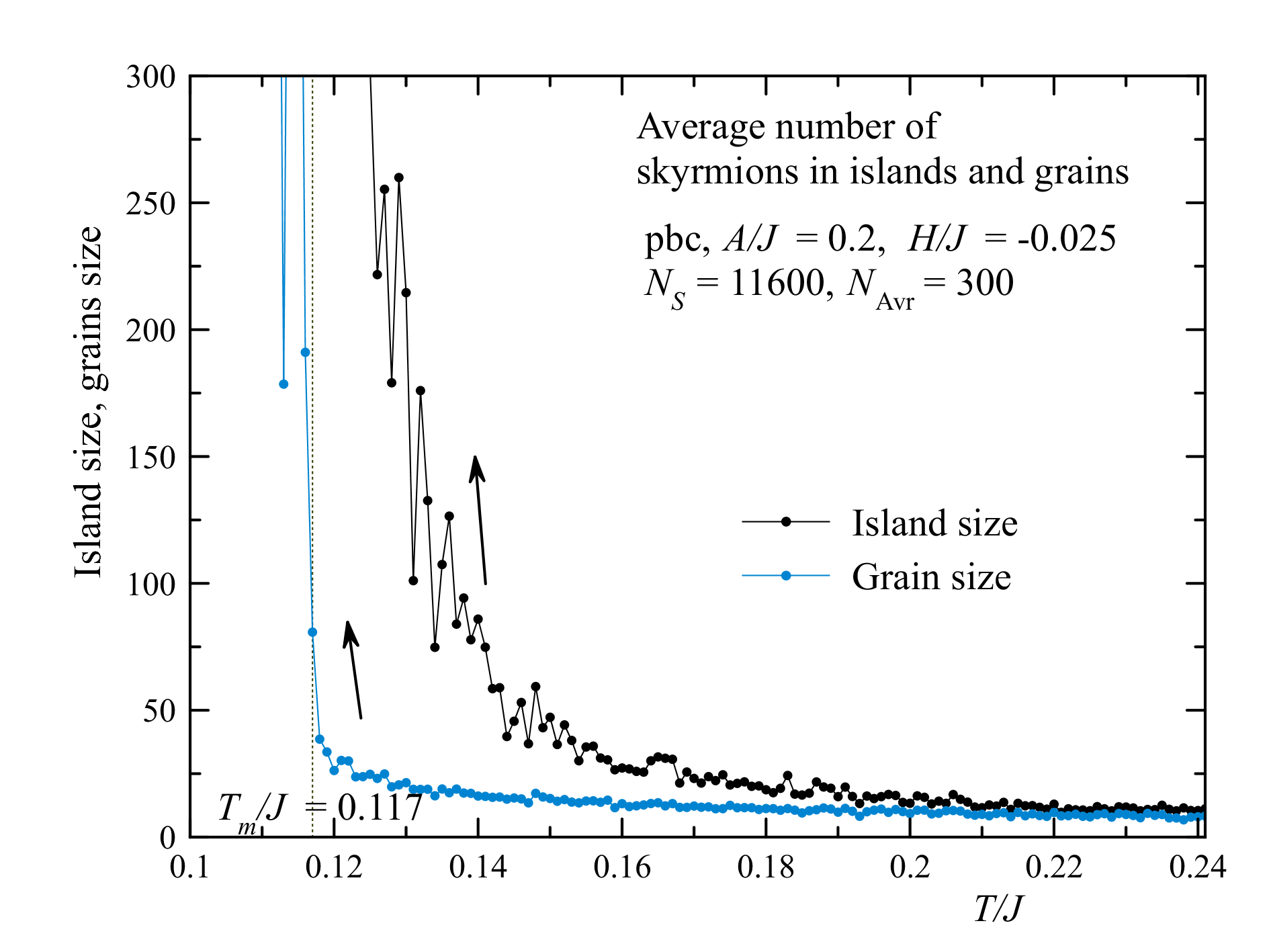}
\par\end{centering}
\caption{Temperature dependence of the number of islands and grains (upper
panel) and of the average size of islands and grains (lower panel).}

\label{Fig-Number_of_Islands}
\end{figure}

Fig. \ref{Fig-cDisl_vs_T} shows the temperature dependence of the
concentration of lattice defects: dislocations, free dislocations,
and free disclinations. Although dislocations are defined via the
Burgers vector, at nonzero temperatures, when the lattice is deformed,
it is more convenient to look at the Delaunay graph showing the number
of nearest neighbors for each particle. Particles having five of seven
nearest neighbors are disclinations. We count free disclinations which
are separated by a distance of least $2a_{S}$ from other defects.
A pair of 5- and 7-disclinations separated by the distance of order
$a_{S}$ form a dislocation. Again, we count free dislocations which
are separated by a distance of at least $2a_{S}$ from other defects.
Also we count all dislocations, including clustered ones. As can be
seen, only a small fraction of lattice defects are free while the
majority of them are clustered. Moreover, the number of free dislocations
and disclinations decreases with temperature above $T_{m}$. This
happens because the total number of dislocations increases and it
becomes difficult to find a place not close to any of them.

As the temperature increases beyond the melting temperature, the regions
between the grains, where the lattice is broken, become wider. Here
the skyrmion matter become liquid, and the fraction of the liquid
phase increases with temperature. This can be quantified as follows.
The temperature dependence $V_{6}(T)$ shown in the lower panel of
Fig. \ref{Fig-O6_vs_T} suggests using the criterion $V_{6}^{2}>0.5$
for solid and $V_{6}^{2}<0.5$ for liquid. Around the melting/freezing
transition, there are regions of both types, and by computing their
areas (or the numbers of skyrmions in them) one can define the crystalline
fraction. This quantity is shown in red in Fig. \ref{Fig-CrystFrac}.
One can see that there is some crystalline fraction above melting
as well as some liquid fraction below melting. However, besides slow
processes of grain dynamics, there is a fast quasi-harmonic motion
of the particles that creates lattice deformations decreasing the
hexagonality $V_{6}$. To make a more reliable separation between
liquid and solid parts of the system, one can perform a short averaging
of skyrmions' positions over the Monte Carlo process -- so short
that it averages out the quasiharmonic motion but does not affect
the slow grain dynamics. The corresponding results are shown in Fig.
\ref{Fig-CrystFrac} for several different values of the averaging
Monte Carlo steps $N_{avr}$ performed after the system was thermalized.
As the result of positional averaging, the liquid part below $T_{m}$
disappears completely that confirms the hypothesis about its origin
as fully due to the quasi-harmonic motion. The crystalline part above
melting increases as well for the same reason. However, too long averaging
begins to reduce the crystalline part at higher temperatures because
here grain dynamics becomes faster. For revealing grains in the melt,
there is an optimum length of the averaging cycle $N_{avr}=100-300$
MCS, as can be seen in Fig. \ref{Fig-CrystFrac}.

One can enumerate the islands of solid within the liquid with the
help of the connected-component labeling algorithm. In this way, one
can also find the average island size (the average number of skyrmions
in the island). This algorithm can be generalized making it resolved
with respect to the orientation of hexagons. In the following we call
regions of solid with an unresolved hexagon orientation ``island''
and those with a resolved orientation ``grains''. One island can
consist of several grains, so there are more grains than islands.
Some grains are separated from each other by dislocations-rich grain
boundaries. There are also continuous changes of grain orientations
that do not lead to creation of dislocations. To define grains in
this situation, we set intervals of $10^{\circ}$ for hexagon orientations.
When the orientation changes by more than 10 degrees, this is considered
as another grain. As the full interval of hexagon orientations is
$60^{\circ}$, there are only six different grain orientations that
we take into account here.

Temperature dependences of the number of islands and grains, as well
as that of the average island and grain size, is shown in Fig. \ref{Fig-Number_of_Islands}.
Well above $T_{m}$ everything is liquid and the number of islands
and grains goes to zero. With lowering the temperature, grains are
formed and their number increases. However, with a further lowering
temperature, islands begin to coalesce, and their number decreases
after reaching a maximum. In this region, the number of grains still
increases as the big coalesced islands consist of many grains with
different orientations. Finally, at the freezing point all grains
acquire the same orientation and become one grain, that is, a monocrystal.
Below freezing, there is one island and one grain. The size of islands
and grains in the lower panel of Fig. \ref{Fig-Number_of_Islands}
monotonically increases with lowering temperature starting from its
minimal value of six particles in the island or grain. Below $T_{m}$
the island size and the grain size are equal to the number of skyrmions
in the system. 

It should be noted that the state with grains of solid within the
melt is not a mixed state because grains are volatile and there is
no second temperature, above the melting/freezing temperature, at
which grains completely disappear. In most of the temperature range,
except for a close vicinity of $T_{m}$, grains are very small to
be considered as a thermodynamically stable phase. Rather, grains
are correlated regions above the ordering transition, precursors of
the emerging ordering. The average grain size gives an estimation
of the correlation length above the transition.

\section{Nonequilibrium effects in the polycrystalline state}

\label{Sec_Nonequilbrium}

\begin{figure}
\begin{centering}
\includegraphics[width=8cm]{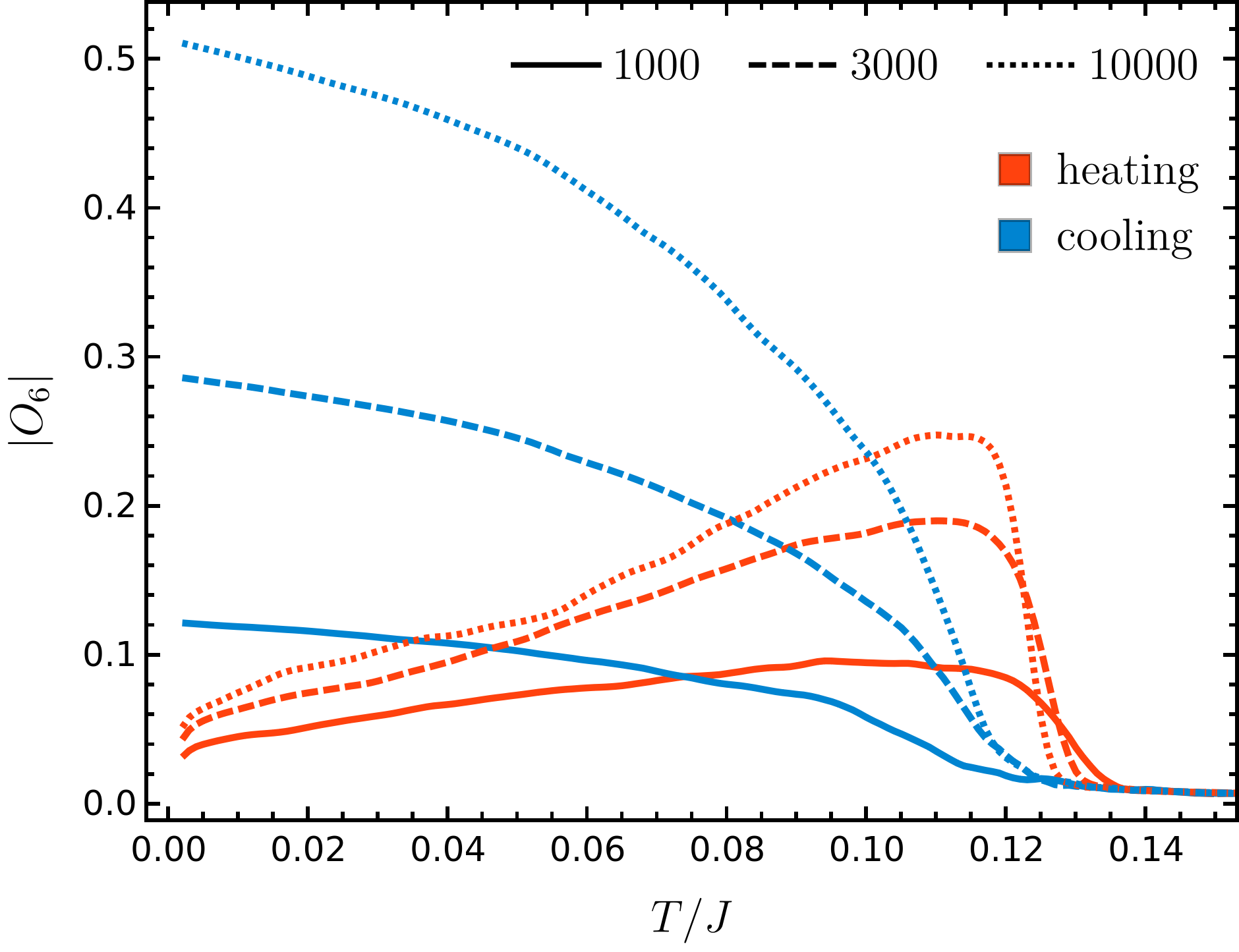}
\par\end{centering}
\centering{}\includegraphics[width=8cm]{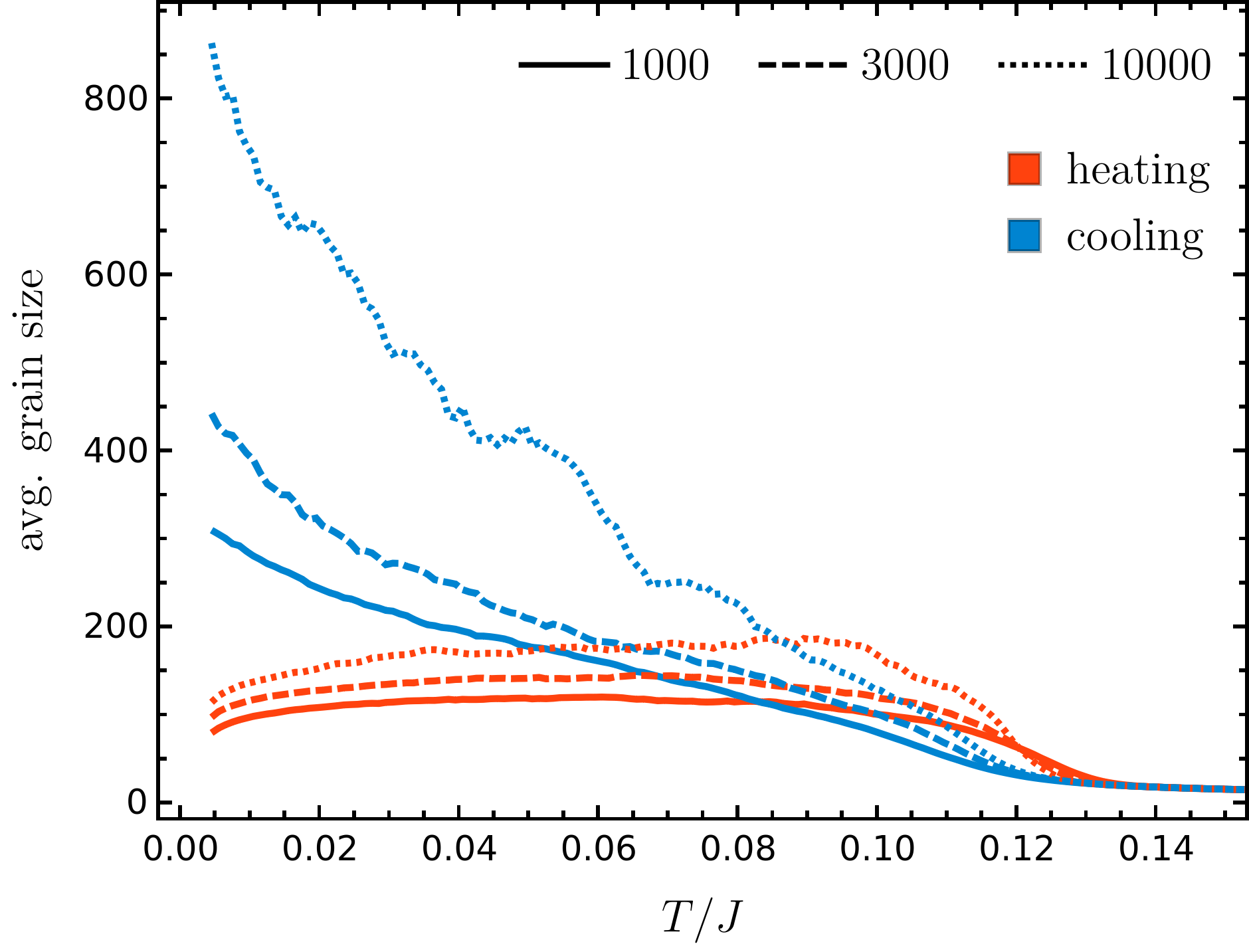}\caption{Nonequilibrium effects on heating and cooling of the skyrmion lattice
(with the initial random placement of skyrmions) have some similarity
to field-cooled and zero-field-cooled phenomena observed in arrays
of magnetic particles. The data was obtained on the system of 104400
skyrmions by stepwise increasing and decreasing $T$ with $\delta T/J=0.001$
and the number of MCS at each temperature 1000, 3000, and 10000. Upper
panel: the orientational order parameter is reduced because the state
is polycrystalline. Lower panel: the average grain size (the number
of skyrmions in the grain) in the polycrystalline state.}
\label{Fig-T-sweeps}
\end{figure}
\begin{figure}
\begin{centering}
\includegraphics[width=8cm]{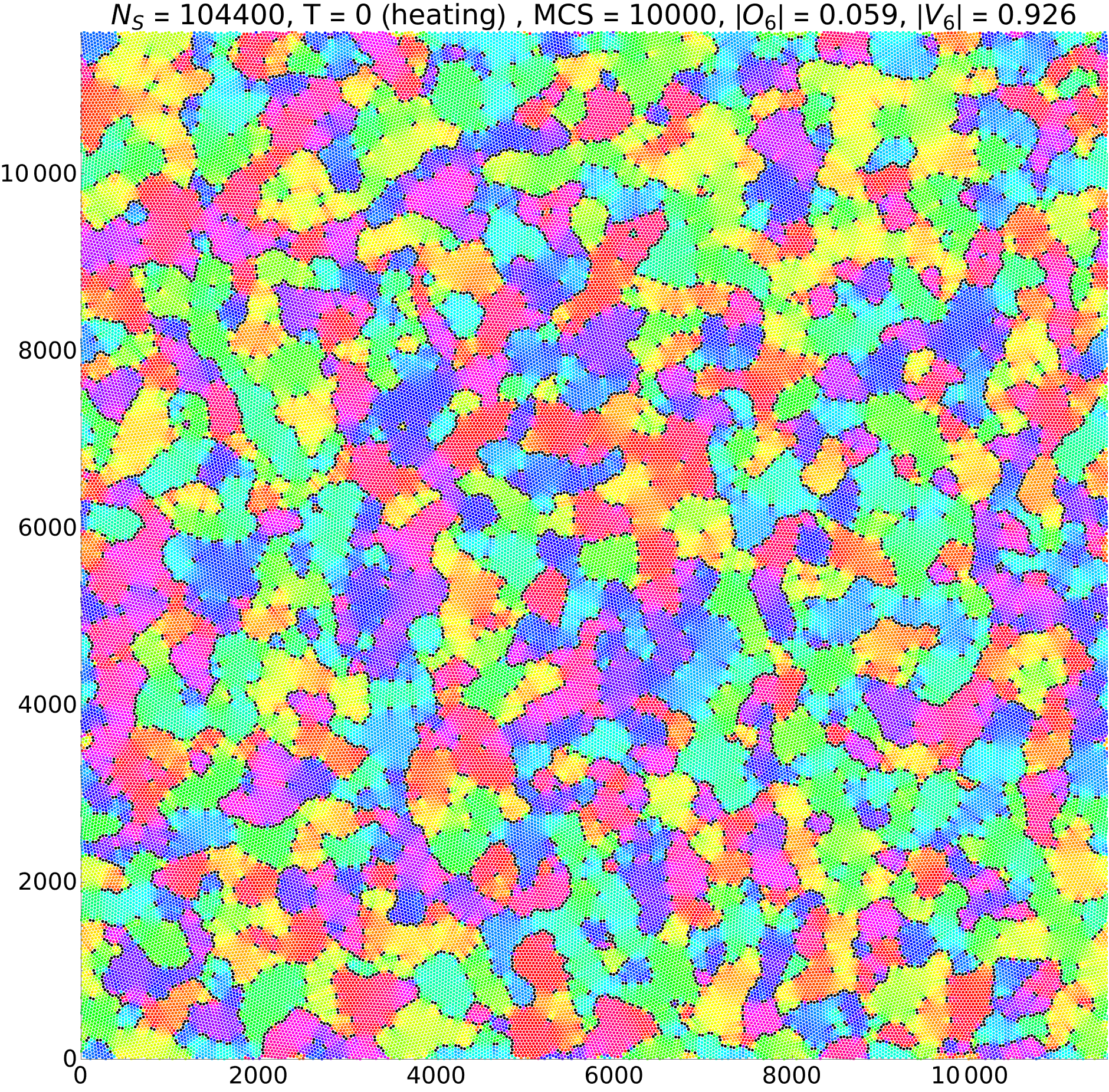}
\par\end{centering}
\begin{centering}
\includegraphics[width=8cm]{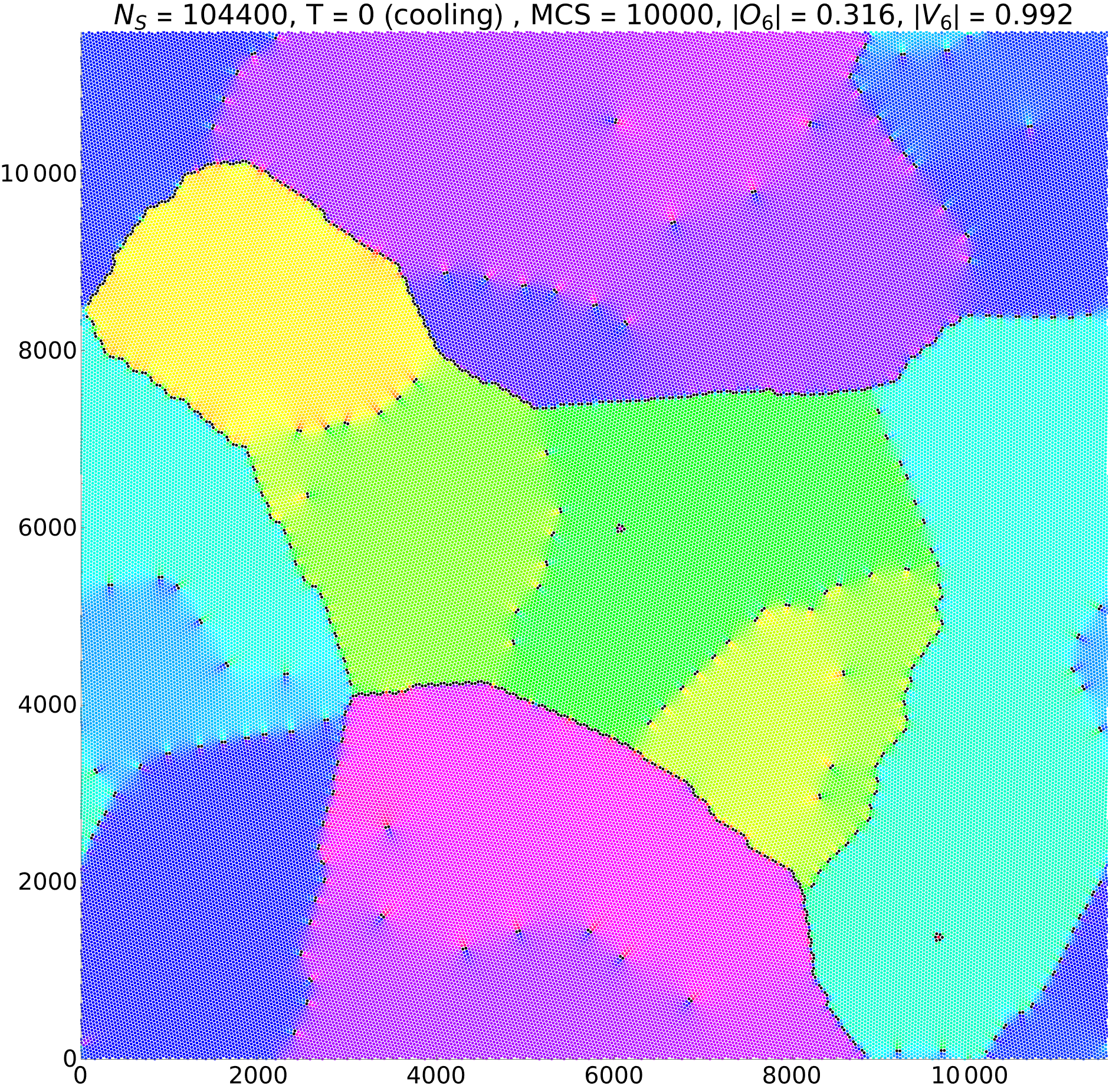}
\par\end{centering}
\caption{The polycrystalline state in the system of 104400 skyrmions at $T=0$
with 10000 MCS per temperature point. The hexagon orientation is color-coded
as above. Skyrmions having 5 or 7 Delaunay nearest neighbors are shown
in black. Upper panel: small grains obtained by quenching from the
initial state with a random placement of skyrmions. Such states are
then warmed up that leads to coarsening of grains. Lower panel: Large
grains obtained by gradual lowering the temperature.}

\label{Fig-Grains_at_T=00003D0}
\end{figure}

Large systems of skyrmions on lowering the temperature freeze into
a polycrystalline state. The pure system with periodic boundary conditions
that we are studying here then tends to evolve very slowly towards
the monocrystalline state via the motion of boundaries between differently
oriented grains. If there is only one flat boundary between the two
large grains, the system will stay in this state forever. Also even
a weak pinning or the effect of boundaries will prevent the system
from reaching the monocrystalline state. In particular, a square system
with rigid boundaries is a frustrated system because differently oriented
rigid boundaries favor different orientations of hexagons.

To elucidate the evolution of the polycrystalline state, we performed
stimulated annealing experiments on the large system with 104400 skyrmions
by stepwise increasing and decreasing the temperature with the $T$-step
$\delta T/J=0.001$ and the number of MCS at each temperature 1000,
3000, and 10000. In the last case, the total number of MCS done in
changing the temperature between zero and $T/J=0.1$ is $10000\times0.1/0.001=10^{6}$.
Both for warming and cooling, we start with a random placement of
skyrmions. At $T=0$, the system relaxes into a metastable state with
small differently oriented grains and thus a small value of the orientational
order parameter $O_{6}$. The process of grains coarsening is very
slow at low temperatures. With increasing $T$, the coarsening goes
faster and $\left|O_{6}\right|$ increases up to the point of melting
before it finally goes down. This behavior has a similarity to that
observed in the zero-field-cooled experiments on magnetic systems.
However, there is no analog of the applied field that causes increasing
of the magnetization with increasing $T$, and $\left|O_{6}\right|\propto\sqrt{\mathrm{GrainSize}/N_{S}}$
decreases with the system's size. 

A more suitable quantity characterizing the polycrystalline state
is the average grain size, as defined in the preceding section. This
quantity does not depend on the system size for large systems. As
the grain size can be rather big, studying the polycrystalline state
requires large systems. Fig. \ref{Fig-T-sweeps} shows the temperature
dependences of $\left|O_{6}\right|$ and the grain size in the system
of 104400 skyrmions. With increasing the number of MCS done at each
$T$-point, both $\left|O_{6}\right|$ and the grain size increase,
although much more MCS is needed to reach the monocrystalline state
with $\left|O_{6}\right|=1$ in the limit $T\rightarrow0$ and the
grain size equal to the system size. 

Examples of the states of the system of $N_{S}=104400$ with 10000
MCS per temperature point at $T=0$ are shown in Fig. \ref{Fig-Grains_at_T=00003D0}.
The upper panel shows small grains obtained by quenching from the
initial state with a random placement of skyrmions. Such states are
then warmed up that leads to the coarsening of grains, see warming
curves in Fig. \ref{Fig-T-sweeps}. The lower panel shows large grains
obtained by gradual lowering the temperature, see cooling curves in
Fig. \ref{Fig-T-sweeps}.

\section{Conclusions}

\label{Sec_Conclusions}

Skyrmion lattices provide another field for testing the predictions
of the theory of 2D melting. They present an advantage over systems
of particles due to the possibility of continuously changing interaction
between skyrmions by changing the external magnetic field. Here we
performed comprehensive Monte-Carlo studies of the melting of 2D skyrmion
lattice. The temperature and the field dependence of the positional
and orientational order parameters have been computed in systems of
various sizes ranging from $10^{3}$ to $10^{5}$ skyrmions, and images
of skyrmion lattices near the melting transition have been obtained.
Skyrmions repel each other with a force that goes down exponentially
with the distance between the skyrmions. The decay length is determined
by the magnetic field. When the system is small enough, thermal equilibrium
is achieved, and we observe a sharp reversible solid-liquid transition
without an intermediate hexatic phase. Larger systems fail to achieve
thermal equilibrium and exhibit hysteresis on cycling the temperature.
For both equilibrium and non-equilibrium systems we find that the
melting and freezing transitions occur via the formation of grains
with different orientations of hexagonal axes. The departure from
a simple picture of a transition caused by the unbinding of dislocation
and disclination pairs is due to the clustering of the defects into
the grain boundaries. We observe the granular structure on both heating
and cooling, with the only difference that in the liquid phase it
corresponds to the grains of a fluctuating shape while in the solid
phase it corresponds to a more stable polycrystal. On cooling the
fluctuating grains freeze into a polycrystal which behaves similarly
to ordinary crystals on annealing. Our findings show that the scenarios
of melting driven by the unbinding of free defects and melting driven
by the formation of grains are likely to be the two limiting cases
of a more complex scenario in which many-body interaction between
the defects plays a crucial role. 

\section{Acknowledgments}

This work was supported by Grant No. DE-FG02-93ER45487 funded by the
U.S. Department of Energy, Office of Science, and Grant No. FA9550-24-1-0090
funded by the Air Force Office of Scientific Research.

\end{document}